\documentclass[sigconf,review=false,anonymous=false]{acmart} 
\usepackage{times}
\usepackage{graphicx}
\usepackage{amsmath}
\usepackage{booktabs}
\usepackage{colortbl}
\usepackage{booktabs}
\usepackage{hhline}
\usepackage{subcaption}
\usepackage{tikz}
\usepackage{lipsum}
\usepackage{paralist}
\usepackage{color}
\usepackage{makecell}
\usepackage{tabularx}
\usepackage{utfsym}
\usepackage{textcomp}
\usepackage{svg}
\usepackage{siunitx}
\usepackage{dirtytalk}
\usepackage{multirow}
\usepackage{arydshln}

\usepackage[ruled,vlined]{algorithm2e}
\renewcommand{\KwSty}[1]{\textnormal{\textcolor{blue!90!black}{\ttfamily\bfseries #1}}\unskip}
\newcommand{\var}{\texttt}
\newcommand{\FuncCall}[2]{\texttt{\bfseries #1(#2)}}
\SetKwProg{Function}{def}{:}{}

\usetikzlibrary{calc}

\newcommand{\repeatcommand}[2]{%
  \ifnum#1>0
    #2%
    \repeatcommand{\numexpr#1-1\relax}{#2}%
  \fi
}

\newcommand{\eg}{\emph{e.g.}}

\AtBeginDocument{%
  }

\setcopyright{acmcopyright}
\copyrightyear{2025}
\acmYear{2025}
\acmDOI{XXXXXXX.XXXXXXX}
\renewcommand\footnotetextcopyrightpermission[1]{}
\acmPrice{15.00}
\acmISBN{978-1-4503-XXXX-X/18/06}

\usepackage{xcolor}

\begin{document}

\title[ObjectFinder for Interactive Object Search]{ObjectFinder: An Open-Vocabulary Assistive System for Interactive Object Search by Blind People
}

\author{Ruiping Liu}
\affiliation{%
  \institution{Karlsruhe Institute of Technology}
  \city{Karlsruhe}
  \country{Germany}
}

\author{Jiaming Zhang}
\authornotemark[1]
\affiliation{%
  \institution{Karlsruhe Institute of Technology}
  \city{Karlsruhe}
  \country{Germany}
}

\author{Angela Sch\"on}
\affiliation{%
  \institution{Karlsruhe Institute of Technology}
  \city{Karlsruhe}
  \country{Germany}
}

\author{Karin M\"uller}
\affiliation{%
  \institution{Karlsruhe Institute of Technology}
  \city{Karlsruhe}
  \country{Germany}
}

\author{Junwei Zheng}
\affiliation{%
  \institution{Karlsruhe Institute of Technology}
  \city{Karlsruhe}
  \country{Germany}
}

\author{Kailun Yang}
\affiliation{%
  \institution{Hunan University}
  \city{Changsha}
  \country{China}
}

\author{Anhong Guo}
\affiliation{%
  \institution{University of Michigan}
  \city{Ann Arbor, MI}
  \country{USA}
}

\author{Kathrin Gerling}
\affiliation{%
  \institution{Karlsruhe Institute of Technology}
  \city{Karlsruhe}
  \country{Germany}
}

\author{Rainer Stiefelhagen}
\affiliation{%
  \institution{Karlsruhe Institute of Technology}
  \city{Karlsruhe}
  \country{Germany}
}
\authorsaddresses{Corresponding author: Jiaming Zhang,
\href{mailto:jiaming.zhang@kit.edu}{jiaming.zhang@kit.edu};
Karlsruhe Institute of Technology, Karlsruhe, Germany}
\renewcommand{\shortauthors}{Liu, et al.}

\begin{teaserfigure}
    \centering
    \includegraphics[width=\textwidth]{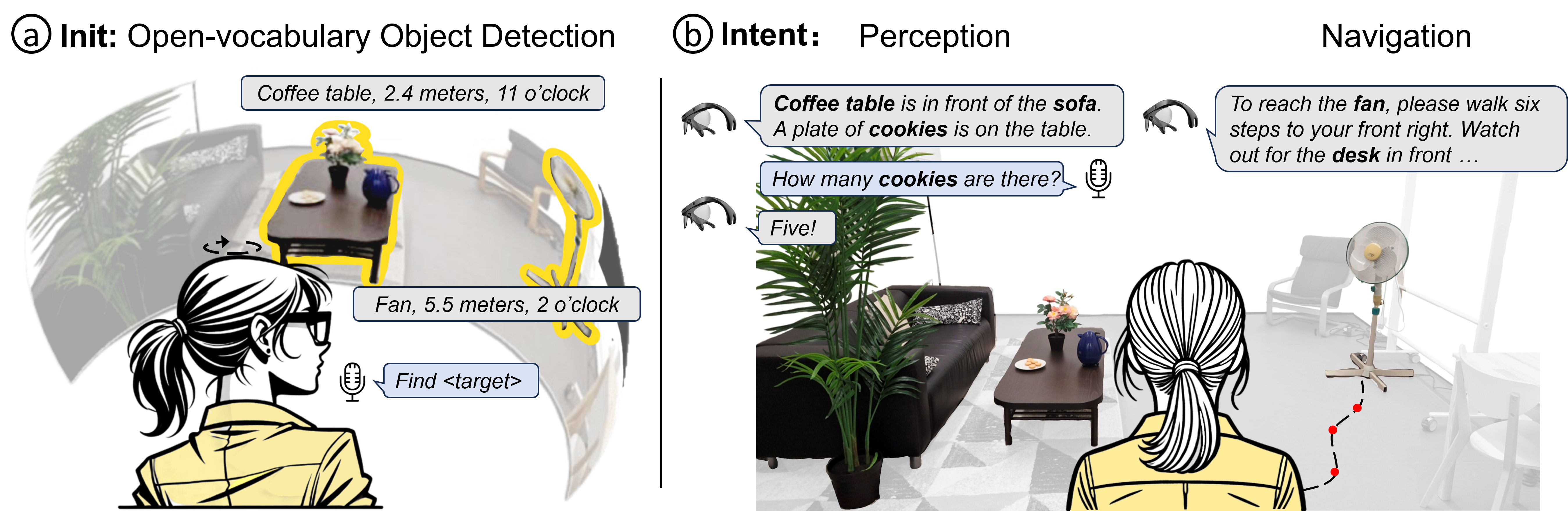}
    \caption{
    ObjectFinder system for open-vocabulary interactive object search. It seamlessly integrates open-vocabulary models, \textit{i.e.} an open-vocabulary object detector (\textit{e.g.}, YOLO-World) and a multimodal large language model (\textit{e.g.}, GPT-4). (a) A user specifies a target with flexible wording on smart glasses. Once it is found, the user is informed with egocentric localization information in real-time. 
    (b) Upon detecting the target object, the user may have various intentions towards it, such as perceiving the top of the coffee table or navigating towards a fan to turn it on. During the interaction, the user may discover other objects of interest for subsequent searches, \eg~cookies on the coffee table.
    }
    \Description{Figure consisting of two panels. On the left, titled 'Init: Open-vocabulary object detection', a blind person using smart glasses asks 'Find <target>' while scanning the area. Detected objects are highlighted; the system relays real-time locations like 'coffee table, 2.4 meters, 11 o'clock' and 'fan, 5.5 meters, 2 o'clock'. The right panel depicts the blind person with various intents for the objects. For the coffee table, the system describes 'It's in front of the sofa with a plate of cookies on top.' Upon hearing about the cookies, she inquires, 'How many cookies are there?', to which the system responds, 'Five'. Interested in the fan, she initiates route planning, and the system guides, 'Walk six steps to your front right, watch out for the desk ahead...'}
    \label{fig:teaser}
\end{teaserfigure}

\begin{abstract}
Searching for objects in unfamiliar scenarios is a challenging task for blind people. It involves specifying the target object, detecting it, and then gathering detailed information according to the user's intent. 
However, existing description- and detection-based assistive technologies do not sufficiently support the multifaceted nature of interactive object search tasks.
We present ObjectFinder, an open-vocabulary wearable assistive system for interactive object search by blind people. ObjectFinder allows users to query target objects using flexible wording. Once the target object is detected, it provides egocentric localization information in real-time, including distance and direction. Users can then initiate different branches to gather detailed information based on their intent towards the target object, such as navigating to it or perceiving its surroundings. ObjectFinder is powered by a seamless combination of open-vocabulary models, namely an open-vocabulary object detector and a multimodal large language model. The ObjectFinder design concept and its development were carried out in collaboration with a blind co-designer. 
To evaluate ObjectFinder, we conducted an exploratory user study with eight blind participants. We compared ObjectFinder to BeMyAI and Google Lookout, popular description- and detection-based assistive applications. Our findings indicate that most participants felt more independent with ObjectFinder and preferred it for object search, as it enhanced scene context gathering and navigation, and allowed for active target identification. Finally, we discuss the implications for future assistive systems to support interactive object search.

\end{abstract}

\maketitle

\section{Introduction}
\begin{table*}[t]
    \centering
    \caption{Comparison of different systems that can be used for object search.}
    \resizebox{\textwidth}{!}{%
    \begin{tabular}{lcccc}
    \toprule
         \textbf{System} & \textbf{Purpose} & \textbf{Enabling Source} & \textbf{Interaction} & \textbf{Device}\\
    \midrule
         RSA~\cite{bemyeyes, taptapsee, aira}& Multi-Purpose& Human& Dialogue&Smartphone\\
         ProgramAlly~\cite{herskovitz2024programally}& Object Search& AI& Filter Customization&Smartphone\\
         WorldScribe~\cite{chang2024worldscribe}& Exploration& AI& Intent Customization& Smartphone\\
         WanderGuide~\cite{kuribayashi2025wanderguide}& Exploration& AI& Dialogue, Button-Driven Option Selection& Robot\\
         BeMyAI~\cite{BeMyAI2023}& Description& AI& Dialogue& Smartphone\\
         Lookout~\cite{lookout}& Exploration& AI& -& Smartphone\\
         Find My Things~\cite{wen2024findmythings}& Object Search& AI& Teachable Object Recognition& Smartphone\\
         LifeInsight~\cite{mathis2025lifeinsight}&Question and Answering& AI& Dialogue& Wearable Device\\
    \midrule
         ObjectFinder& Object Search \& Exploration & AI& Dialogue, Button-Driven Option Selection& Wearable Device\\
    \arrayrulecolor{black}
    \bottomrule
    \end{tabular}
    }
    \label{tab:techs}
    
\end{table*}
Blind people often face challenges when searching for objects in unfamiliar environments~\cite{muller2022traveling, zhang2021trans4trans}. Independently searching for a specific object, such as locating the nearest available chair in a spacious lobby through haptic exploration, can be particularly difficult. To search for an object, users would first query the target object, and detect a candidate.
In order to determine if it is the desired one, both egocentric (\textit{e.g. } ``\textit{11 o'clock, 2.4 meters away}'') and allocentric (\textit{e.g.} ``\textit{next to the desk}'') information are necessary for a blind user to perceive the object in their environment~\cite{golledge1999wayfinding,martolini2021allocentric}. 
Upon locating the target object, the user's intent may vary. 
As shown in examples of Figure~\ref{fig:teaser}, 
once the coffee table is detected, the user might prefer an audio description of items on the table over immediate physical interaction. Conversely, if the target is a fan, the user might wish to navigate towards it to turn it on. Throughout the search, there may also be a desire to explore the surroundings for better navigation and potentially discover other targets for subsequent searches~\cite{jain2023want}.

Such an interactive object search task is multifaced, however, no current AI-powered assistive technology can yet handle all the associated subtasks. 
We categorize the existing assistive technologies for blind people into 
\emph{description-based} and \emph{detection-based} systems. 
Description-based systems provide vivid and detailed descriptions of a photo~\cite{BeMyAI2023} or brief captions~\cite{seeingAI,lee2022imageexplorer,liu2023opensu}. 
However, these systems are unable to localize a specific object in an unfamiliar environment while ensuring that the target object is in the frame~\cite{xie2024emerging} 
\textbf{(Challenge 1, C1)}. Detection-based systems~\cite{lookout, ai_poly, zheng2023materobot}, on the other hand, either allow only the search for 
a limited number of pre-defined objects~\cite{constantinescu2020blind, Yi2013, schauerte2012assistive, ahmetovic2020recog, kacorri2017classifier, lookout, ai_poly, wen2024findmythings} or provide filtered information~\cite{herskovitz2024programally, chang2024worldscribe}, limiting understanding of unfamiliar scenes. 
Therefore, when using current detection-based systems in an unfamiliar environment, blind users may not know what is in a room comprehensively and miss 
items that could be of interest 
\textbf{(Challenge 2, C2)}. Apart from that, both detection-based and description-based systems fail to provide egocentric information (distance and direction) or support question and answering directed towards the target.

The challenges also exist in remote sighted assistance (RSA). The procedure by which the remote agents~\cite{aira, bemyeyes,taptapsee} help to identify objects and describe surroundings involves capturing images from the video feed and zooming in to obtain the necessary visual information~\cite{lee2020emerging}. In this context, it is time-consuming for remote agents to adjust the video frame, and they find it challenging to continuously orient the users~\cite{xie2022iterative}. Moreover, recognizing landmarks presents significant difficulties for the agents~\cite{lee2022opportunity, xie2022help, kamikubo2020support}. Thus in this work, we aim to address the following question:

\textit{How to integrate the advantages of description- and detection-based assistive systems to support interactive object search by blind people?}

To this end, we designed 
ObjectFinder, which seamlessly combines open-vocabulary models, an Open-Vocabulary Object Detector (OVOD)\cite{Cheng2024YOLOWorld} and a Multimodal Large Language Model (MLLM)\cite{openai-gpt4}, to facilitate an interactive process that ranges from object detection to description for object search. 
Users can input any target via voice commands for object detection, then scan the scene. Once a candidate is detected, the system will notify the user to stand still to orient to the target and output real-time egocentric information (distance and direction). Following this, users can acquire comprehensive information about their surroundings, tailored to their intent, based on the keyframe captured at the time of detection. This process facilitates the identification of potentially interesting and unexpected targets, which can then be explored in further detail during subsequent iterations.

Based on prior works in object search tasks, we formulated design considerations for an interactive system that enables flexible querying, supports various search subtasks, and adapts the system feedback to user intent.
We co-designed with a blind person throughout the conception and development of ObjectFinder, across two months and four iterations. We deconstruct the complex object search task into the following functions: target object detection, localization, route planning, scene description, and open questions. The pipeline for integrating all functions and interaction features was refined based on the iterative feedback from the blind co-designer. 

To evaluate the effectiveness and efficiency of ObjectFinder in facilitating object search, we conducted an exploratory evaluation with eight blind participants. They engaged with the system and participated in a semi-structured interview afterwards.
BeMyAI~\cite{BeMyAI2023} and Google Lookout~\cite{lookout}, popular commercial description- and detection-based systems, were used as baseline comparisons.
Through thematic analysis~\cite{FeredayMuirCochrane2006}, we demonstrate that the use of ObjectFinder enhanced interactive object search. It integrated crucial information about egocentric localization (distance and direction) from the detection component, and allocentric relationships among objects from the description component. 
Additionally, route planning was a valuable feature of ObjectFinder for searching objects. 
Although ObjectFinder provides feedback based on users’ intents, individual variations in procedures and information preferences, also influenced by a scenario’s scope and familiarity, underscore the need for customization and personalization, as discussed later. ObjectFinder represents a significant advancement in bridging the technological gap in object search, particularly in unfamiliar contexts. Taking advantage of both description- and detection-based systems, its technical approach has the potential to find systems broadly to enhance the independence of blind individuals in their daily lives.

\section{Related Work}
In this section, we introduce the task of object search for blind individuals and provide an overview of existing description-based and detection-based systems designed specifically for them, which can partially address the task. Since no assistive system currently exists for object search in unfamiliar scenarios, we refer to procedures from embodied AI for object search, which typically mimic human behavior. This forms the background knowledge for our study.
\subsection{Object Search in Unfamiliar Scenarios}
Object search is a multifaceted task that involves object detection, exploration, navigation, and more. 
In addition to small items that blind individuals frequently search for in their daily lives such as smartphone, keys and wallet~\cite{netz2025what}, they often search for large, salient objects as landmarks to improve orientation in unfamiliar environments~\cite{ATMAPS-D2.1, yang2010context}. When searching for objects in unfamiliar environments, blind people typically seek an initial overview of the space, followed by specific details as required~\cite{chang2024worldscribe, shneiderman1996eyes}.
If the target object has been found, blind people may have various intentions regarding it. 
For example, they might navigate to the object to interact with it~\cite{herskovitz2023diy} (\eg, find a free chair and sit on it), identify a specific object~\cite{brady2013visual, hong2024understand} (\eg, check whether a bottle is shampoo), or perceive the surroundings of the object (\eg, the tabletop~\cite{herskovitz2023diy}), which may be too far or inconvenient to touch~\cite{Penuela2024usecase}. 
A participant in~\cite{herskovitz2023diy} defined the use case of locating an empty chair in the classroom and imagined how the object search system should work: he preferred to scan the environment with smart glasses rather than waving his phone in the crowd, then the system find an empty chair and give directions on how to walk to the chair. 

Some technologies have been proposed to partially address the challenges of object search (Table~\ref{tab:techs}). Vizwiz-LocateIt~\cite{bigham2010locateit} lets users photograph target objects, ask questions to a remote worker on Mechanical Turk, and navigate via sonification. 
Tools such as AIRA~\cite{aira}, Vizwiz~\cite{gurari2018vizwiz}, and BeMyEyes~\cite{bemyeyes} utilize crowdsourcing to connect blind people with sighted people for real-time remote sighted assistance including object search. However, asking the blind people to move their phone to adjust the video frame is time-consuming~\cite{lee2020emerging}. 
WanderGuide~\cite{kuribayashi2025wanderguide} has subfunctions for object search implemented on a suitcase, but is designed primarily for exploration without specific consideration of the object search procedure.
Bhanuka \textit{et al.}~\cite{bhanuka2023want} suggest that the conversational interface on wearable devices is suitable for the complex task of providing environmental information. We categorize existing AI-based assistive technology related to object search into description-based and detection-based systems.

\subsection{Description-based Systems for Blind People}
The description-based systems capture the scene and describe it only once. 
Seeing AI~\cite{seeingAI}, ImageExplorer~\cite{lee2022imageexplorer}, and OpenSU~\cite{liu2023opensu} generate brief image captions for the scenes captured by a mobile phone and enable tactile exploration of the salient objects on the touchscreen. 
TapTapSee~\cite{taptapsee} is an application that generates a concise phrase about the salient object in almost real-time. 
BeMyAI~\cite{BeMyAI2023}, a feature of BeMyEyes empowered by GPT-4, delivers vivid descriptions of the scenario and allows users to ask questions. Research on BeMyAI~\cite{xie2024emerging,xie2025beyond} indicates that while it serves as a form of distributed cognition, it faces challenges in intent recognition and frequently necessitates the use of multiple images to accurately convey information. NaviGPT~\cite{zhang2025navigpt} is a mobile navigation system that provides a brief description of the road ahead.
LifeInsight~\cite{mathis2025lifeinsight} is a wearable system embedded with GPT-4 for question answering.
Some other works focus on the specific features of the salient object, \textit{e.g.}, material~\cite{zheng2023materobot}, transparency~\cite{zhang2021trans4trans}, and various hazards~\cite{yang2018unifying,yang2018predicting}. 
However, people with blindness should make the object within the region of interest captured by mobile devices while using description-based applications~\cite{yang2024viassist,gurari2018vizwiz}. 
In addition, we found that people with blindness can barely align the photos they capture with real spatial dimensions using mobile phones during our user study, which is in line with the findings of~\cite{kacorri2017classifier}. However, Gonzales~\textit{et al.}~\cite{Penuela2024usecase} determined that the primary goal for users of AI-powered scene description applications is to identify specific objects.
Therefore, we implemented an object detector to identify the region of interest for the description module, ensuring a precise understanding of the area where the target is located.

\subsection{Detection-based Systems for Blind People}
Detection-based systems are designed to provide real-time outputs of identified objects or features of interest. 
Lookout~\cite{lookout} and AIPoly~\cite{ai_poly} exemplify this capability by identifying the nearest object within the phone's field of view. 
Various studies have developed wearable systems~\cite{liu2023realtime, ISLAM2023e16924, Sugashini2024yologlass} with similar functionalities, offering real-time object information through multiple interaction modes. 
WorldScribe~\cite{chang2024worldscribe} delivers real-time descriptions of the current view, tailoring the information based on distance and user intent.
Research has explored the detection of personal objects using methods such as SIFT~\cite{constantinescu2020blind, Yi2013, schauerte2012assistive} and advanced deep learning networks~\cite{ahmetovic2020recog, kacorri2017classifier, wen2024findmythings}. ProgramAlly~\cite{herskovitz2024programally} allows users to customize the information filter and efficiently detect specific features of an object. This suggests that a predefined list is not preferred for exploration.
Navigation assistive systems utilize detection-based methods for obstacle avoidance~\cite{Bala2023, liu2021hida, ou2022indoor}, risk assessment~\cite{wang2024visiongpt}, object finding~\cite{li2023multi, duh2020veye, hu2022stereopilot}, shopping~\cite{boldu2020aisee} and passable path planning~\cite{zou2023realtime_passable, surougi2023realtime_path_planning, hong2024spvinet, jain2023towards}.
These systems autonomously select information, which may limit user agency in actively specifying targets. 
Constantinescu~\textit{et al.}~\cite{constantinescu2020bring} propose a system that allows users to choose from a limited set of objects to receive audible feedback in their vicinity. 
In detection-based systems, the information available to users is limited by system constraints, which prevents them from gaining a comprehensive understanding of essential scene context for object search. 

\subsection{Reference Procedure for Object Search}Since no existing AI system fully addresses the challenges of object search for blind users, we examine the object-search procedures of embodied AI to guide the design of an assistive object search system. Object search is widely recognized as a challenging task that integrates both perceptual and cognitive processes~\cite{sun2025objectnav}. Typically, embodied agents~\cite{aydemir2013active, chaplot2020object, yokoyama2024hm3d} first receive an object query from the user, analyze their surroundings, hypothesize the potential location of the target object, and then plan a navigation path accordingly. Recent workflows leveraging LLMs, such as UniGoal~\cite{yin2025unigoal} and SG-Nav~\cite{yin2024sgnav}, allow robots to continuously explore their environment and match discovered objects with the intended target. CogNav~\cite{cao2024cognav} investigates the modeling of cognitive process of object search, which involves a broad and contextual search back and forth to build a cognitive map. Upon observing the target, it verifies the candidate according to the vicinity, then confirms the candidate. Taioli~\textit{et al.}~\cite{taioli2024collaborative} equipped the object search agent with a self-questioner and an interaction trigger to produce a refined detection description that includes dialogue regarding the target object. 
In this study, we explore the subtasks involved in object search and integrate them into a unified pipeline specifically designed for blind users, taking advantage of both description- and detection-based systems. 

\begin{figure*}[t]
    \centering
    \includegraphics[width=\textwidth]{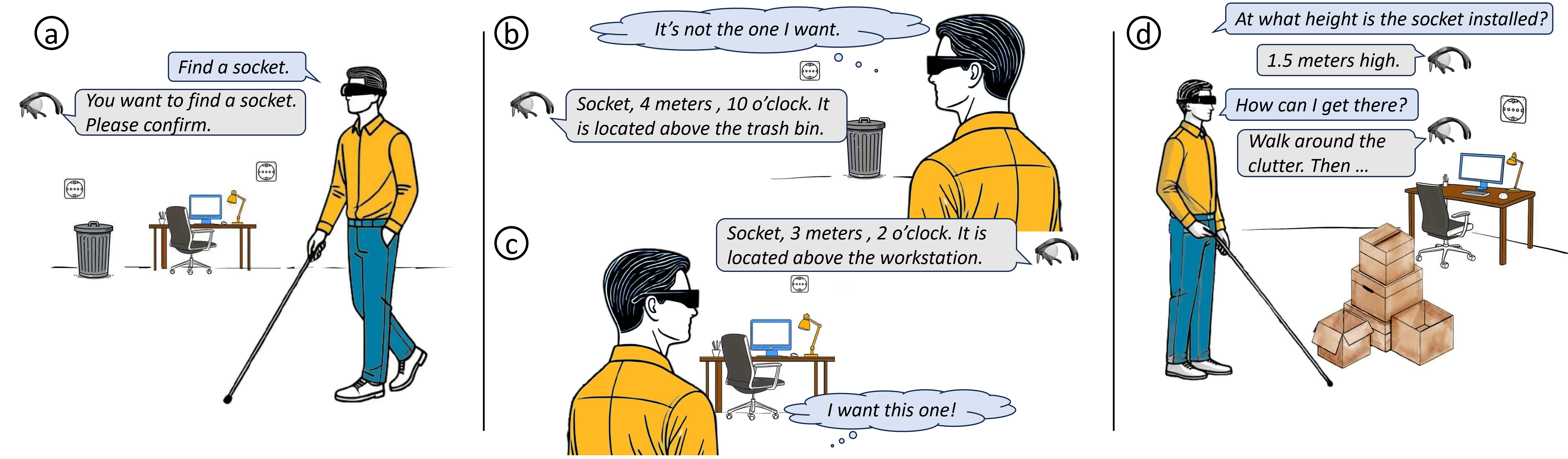}
    \caption{
    Martin walks into an unfamiliar office and uses an object-search system to search for a socket to charge his smartphone. (a) Martin first specifies the target to the system, which then repeats it for confirmation. 
    (b) While scanning, candidates are detected. The socket ``\textit{4 meters away at his 10 o'clock next to the trash bin}'' is not what he wants. (c) However, another socket ``\textit{3 meters away at his 2 o'clock next to the workstation}'' is the desired one, as he plans to study there. (d) After confirming the target, Martin may ask for more details. In large rooms, the system should navigate him to the socket.}
    \Description{Martin walks into an unfamiliar office and uses an object-search system to search for a socket to charge his smartphone. (a) Martin first specifies the target to the system, which then repeats it for confirmation. 
    (b) While scanning, candidates are detected. The socket ``\textit{4 meters away at his 10 o'clock next to the trash bin}'' is not what he wants. (c) However, another socket ``\textit{3 meters away at his 2 o'clock next to the workstation}'' is the desired one, as he plans to study there. (d) After confirming the target, Martin may ask for more details. In large rooms, the system should navigate him to the socket.}
    \label{fig:use_case}
\end{figure*}

\section{ObjectFinder} 
ObjectFinder is a wearable prototype designed for interactive object search. Blind users can specify their target using flexible wording. Once the target is detected, they receive real-time egocentric localization information. They can further obtain detailed feedback based on their intentions toward the targets. We co-designed ObjectFinder with a blind person P0 (see Table~\ref{tab:participants}) by proposing an envisioned scenario, constructing an initial prototype based on it, and then conducting four refinement iterations over two months.
\subsection{Design Goals}
Overall, drawing on related works in object search, we designed ObjectFinder with three primary goals:

\textbf{Providing flexibility in target queries and information retrieval.} Blind people prefer to query flexible target objects and their surroundings, often discovering new items of interest. ObjectFinder should facilitate seamless conversational interactions using open-vocabulary models to bridge wording gaps.

\textbf{Supporting various subtasks during search.} 
Object search is a complex task involving several sequential subtasks: target specification, detection, and feedback generation. ObjectFinder should simplify this process by organizing these subtasks into a user-friendly pipeline with accessible interaction features.

\textbf{Adapting to the user intent of the target object.} 
Blind people exhibit varying intents for identified targets, from navigation to mere perception, and require tailored descriptions or guidance based on the familiarity of their environment. ObjectFinder should offer options that allow users to gather system feedback based on their specific intents.
\begin{figure*}[t]
    \centering
    \includegraphics[width=0.9\textwidth]{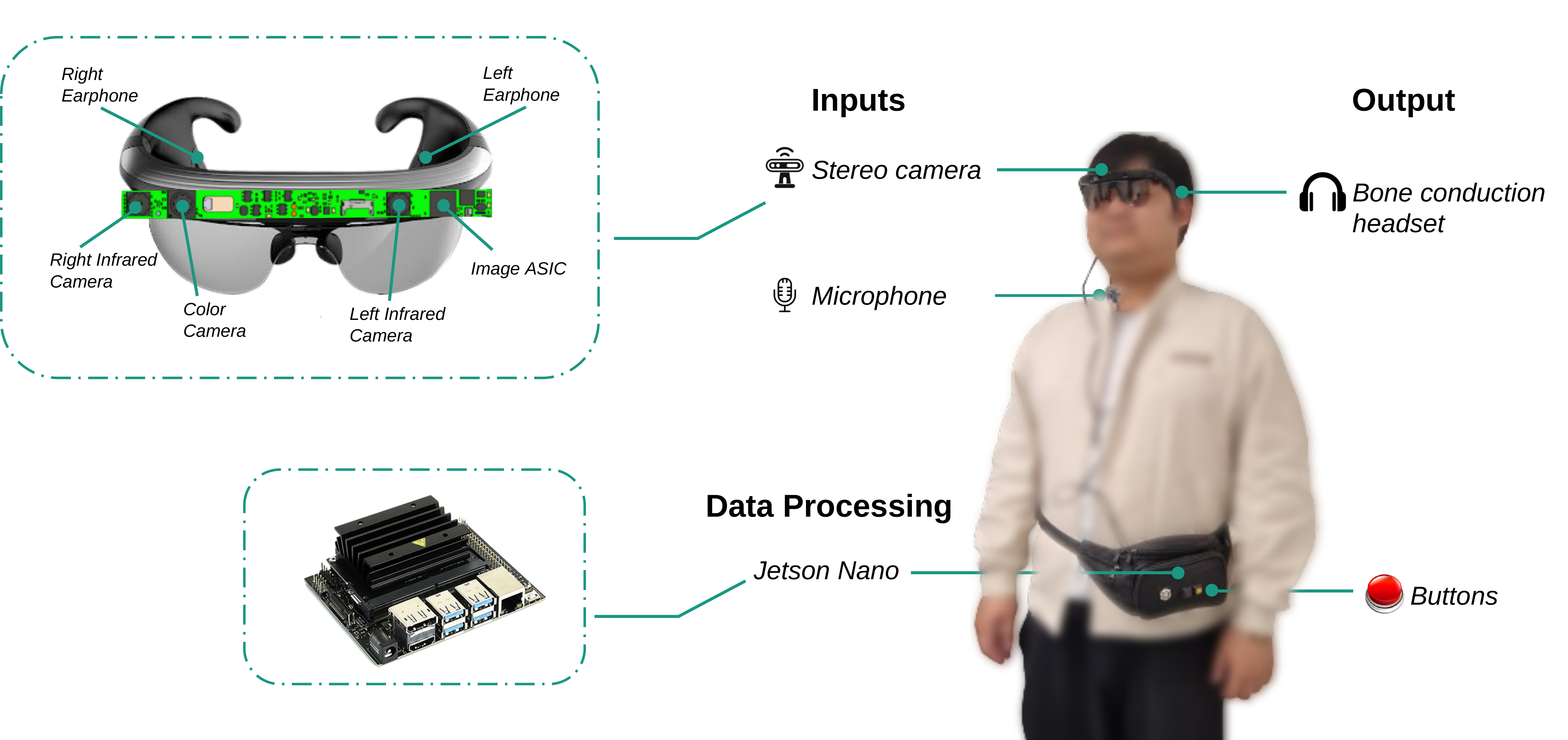}
    \caption{Hardware design and components of the wearable system \textbf{ObjectFinder}. It incorporates a stereo camera to capture visual information about the user's surroundings, a pair of buttons, and a microphone to collect the user's commands. Simultaneously, it executes algorithms through a lightweight processor. To provide a comprehensive and immersive experience, the system delivers spatial-aware informational feedback directly to the user via bone-conduction headphones.}
    \Description{This image displays the components and connectivity of a wearable system. The system consists of smart glasses equipped with a stereo camera, a microphone on the collar, and two buttons on the waist bag for user inputs. The smart glasses also include bone conduction earphones for output. The image further details the internal components of the glasses, which are magnified to show the arrangement from right to left: the right earphone, right infrared camera, color camera, left infrared camera, image ASIC, and left earphone. The NVIDIA Jetson Nano, housed in the waist bag, processes the data collected by these components.}
    \label{fig:hardware}
\end{figure*}
\subsection{Envisioned Scenario}
\label{subsec:use_case} 
In an initial step, we began by defining the use case according to the principles outlined in~\cite{cockburn2000usecase}. To achieve this, we conducted a workshop involving the blind co-designer, a developer, and two experts in accessibility and usability, one of whom is blind.

The co-designer presented a use case for object search: searching for a socket in an unfamiliar office. We refined the use case regarding the interaction sequence between the co-designer and the wearable object search system as the envisioned scenario~\cite{carrol1995scenario}, serves as the basis for our system design, as illustrated in Figure \ref{fig:use_case} and depicted as follows:

Martin enters an unfamiliar office, his phone battery depleted. In need of power, Martin activates an object search system and commands it to \textit{``Find a socket''}. The system acknowledges the command, and ensures Martin that it has understood the request. After Martin confirms, the system signals with a sound, indicating readiness to begin searching.

As Martin scans the room through the system, he prefers not to be bombarded with information about every detected object; instead, he wants the system to announce only when it detects a \textit{socket}. Upon identifying a \textit{socket}, the system provides feedback on its egocentric location, including distance and direction, as well as its allocentric relationship with points of interest.

Using this information, Martin evaluates the suitability of the socket. The first socket detected, located near a trash bin about 4 meters away at a 10 o'clock direction, is deemed inconvenient because Martin intends to work at a workstation. Therefore, he continues his search for a more suitably placed socket.

Eventually, a socket near a workstation, just 3 meters away in the 2 o'clock direction, catches Martin's interest. He requests more details about this socket, such as directions to reach it and its height on the wall. The system advises Martin to navigate around obstacles, guiding him with instructions like,~\textit{``Walk around the clutter...''}

Utilizing his cane to detect and avoid clutter, Martin reaches the workstation situated to his front right and successfully charges his phone using the nearby socket.

\subsection{Hardware and Interaction Features}
\label{subsec:hardware}
According to the related work~\cite{herskovitz2023diy, bhanuka2023want} and the envisioned scenario with the co-designer, a pair of glasses is assumed to be preferred over a smartphone for an object search system. We utilize the following hardware to implement this system.
Figure~\ref{fig:hardware} presents the system diagram, which comprises a pair of KRVision smart glasses~\cite{krvision} coupled with a waist bag. The smart glasses are outfitted with a RealSense R200 RGB-Depth camera, facilitating real-time RGB and depth frame acquisition in an egocentric perspective. Additionally, a bone conduction headset is incorporated, enabling auditory output while maintaining perception of environmental sounds. In the waist bag, an NVIDIA Jetson Nano, a compact and powerful processor, is utilized for efficient data processing, accompanied by a power bank for energy supplementation. 
The waist bag features two buttons that are programmed for target confirmation and function selection.
A microphone is attached to the collar for audio input: participants simply speak commands to specify targets or ask questions after triggering open questions.
The initial version of ObjectFinder is only for the prototype. We aim to further integrate all hardware components more compactly to improve the user's daily experience in real-world use, \eg, with Ray-Ban glasses~\cite{Waisberg2024}.

\subsection{Function Implementation}
\label{subsec:functions}
\begin{figure*}[t]
    \centering
    \includegraphics[width=\textwidth]{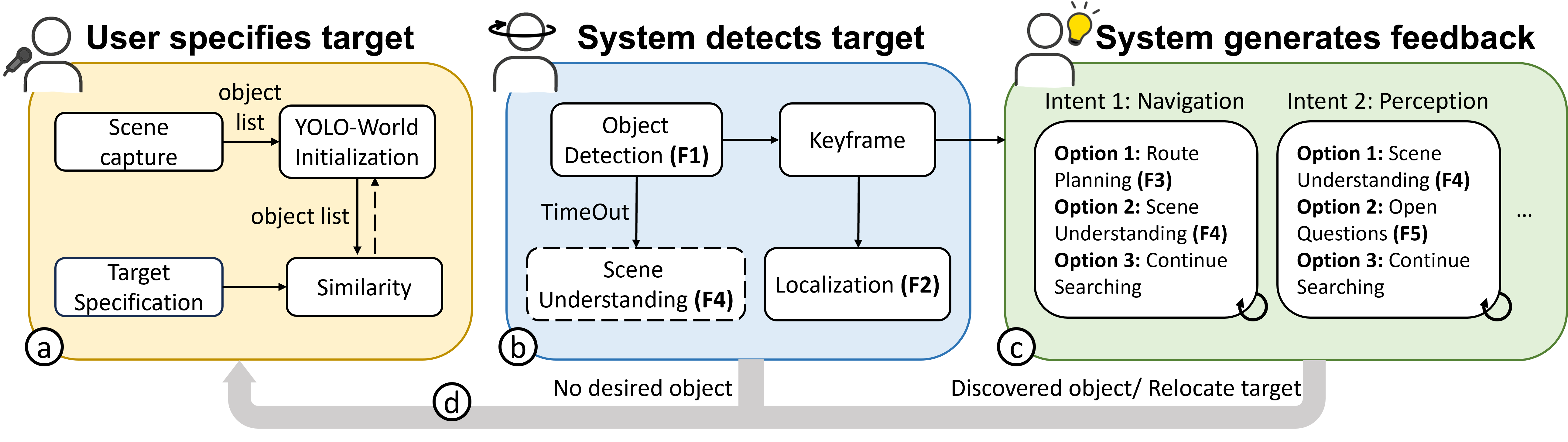}
    \caption{ObjectFinder system architecture integrates five functions into three modules for interactive object search.  (a) Initially, an open-vocabulary object detector, \textit{e.g.} YOLO-World, is initialized with a list of objects extracted from a scenario capture, allowing the user to identify a target object. If the target is not on the list, the object detector is reinitialized. (b) The user scans the environment. If the target is detected, localization information is provided in real-time. If not, the user can trigger scene understanding to identify what exists in the scenario. (c) The user may activate a sub-branch to obtain further information based on their intent using a multimodal large language model. (d) If the user discovers other objects of interest or becomes disoriented, they can reorient themselves to locate the target.}
    \Description{The image shows a system designed to help blind users search for objects. It begins with the user capturing a scene and specifying a target (section a). The system initializes object detection using YOLO-World and compares it with the target. In the perception module (section b), detected objects are localized, or the system moves to scene understanding if detection times out. Based on intent (section c), the system supports navigation (e.g., route planning) or perception tasks (e.g., answering questions or refining the search). Feedback loops allow continuous user-system interaction for effective object search.}
    \label{fig:pipeline}
\end{figure*}

According to the scenario envisioned in Section~\ref{subsec:use_case}, we have decomposed the object search into five functions: Object Detection (F1), Localization (F2), Route Planning (F3), Scene Description (F4), and Open Questions (F5). To integrate these five functions, we define three modules in the pipeline: user specifies the target, system detects the target, and system generates feedback, see Figure~\ref{fig:pipeline}.

\subsubsection{Module 1: User specifies target.}

When the system is turned on, the frame of the scenario is captured by the smart glasses automatically. GPT-4 then generates a list of objects based on the frame to initialize YOLO-World. The user specifies the target object using the command frame, \textit{``Find <target>''}. After receiving the command, the system will repeat the target object: \textit{``You want to find <target>, please confirm.''}. For confirmation, the user should press the button with a sticker on the waist bag, while the other button is for respecification. Speech-to-text is processed by Google Speech Recognition API, and text-to-speech is handled by the pyttsx3.

\begin{figure}[t]
    \centering
    \includegraphics[width=0.9\linewidth]{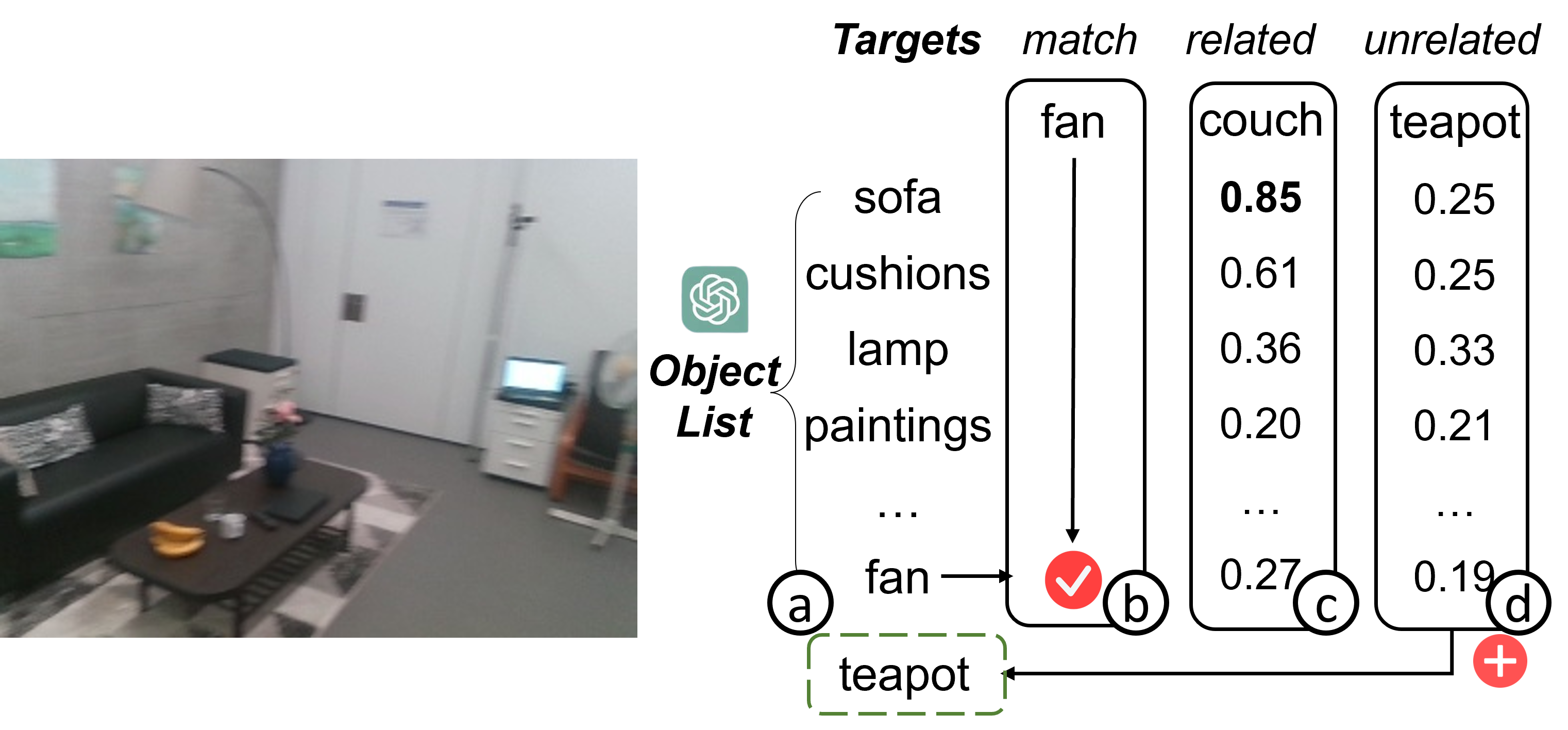}
    \caption{
    Initialization with target identification: (a) The list of detectable objects in YOLO-World is initialized with the first capture of the scenario. The target objects can be categorized into three types: (b) match, where the object matches an item in the list; (c) related, where the object is related to one item in the list (\eg, ``\textit{couch}'' is related to ``\textit{sofa}'' with 0.85 similiarity); and (d) unrelated, where the object does not relate to any item in the list. In cases where the object is unrelated, the list is updated by adding the target to it.}
    \Description{The figure illustrates how a similarity-based matching process works for object search by blind users. On the left, a room scene includes a sofa, cushions, lamp, paintings, and a fan. GPT-4 generates an object list from this scene. Three example target queries are shown: "fan," "couch," and "teapot." The target "fan" directly matches the object list, requiring no similarity calculation. For "couch," the system computes similarities with existing objects: sofa (0.85), cushions (0.61), lamp (0.36), paintings (0.20), and fan (0.27). Since "sofa" has the highest similarity and exceeds the threshold of 0.8, it is selected without needing to reinitialize YOLO-World. In contrast, "teapot" has low similarity scores across all objects—sofa (0.25), cushions (0.25), lamp (0.33), paintings (0.21), and fan (0.19)—none exceeding the threshold. As a result, the system triggers YOLO-World reinitialization using the original object list to find the teapot.}
    \label{fig:target_specify}
\end{figure}

The relationships between the specified target objects and the object classes in YOLO-World are categorized into three types: \textit{match}, \textit{related to}, and \textit{unrelated to}, as shown in Figure~\ref{fig:target_specify}. We define \textit{match} as instances where the string of one item in the object list appears within the target object. This is important because recorded speech may sometimes be unclear due to user rephrasing, such as \textit{``Find the chair, no, office chair.''} and the surrounding conversations.

For other cases, the specified target object and each object class in YOLO-World are tokenized and embedded with all-MiniLM-L6-v2~\cite{reimers2021allMiniLM}. The cosine similarity between the embeddings of the target object and the object classes of YOLO-World is calculated. If the similarity score between the target object and any object class in YOLO-World is at least 0.8, a threshold we have set based on empirical data, the target object is deemed to be \textit{related to} the object class with which it shares the highest similarity in the current list of YOLO-World objects. If the target object is deemed \textit{related to} an existing item in YOLO-World, this item will be used for subsequent object detection searches, thereby eliminating the need to reinitialize the system.
Conversely, if no similarity between the objects in the list and the target object exceeds the threshold, then the target object is considered \textit{ unrelated to} the current object list of YOLO-World. In this case, YOLO-World should be reinitialized with the object list updated to include the new target object. During the YOLO-World initialization process, a $3$ Hz beep is played in the background to reassure the user that the system is still operating.
\subsubsection{Module 2: System detects target.}

After YOLO-World is initialized with the target object, the user will hear an earcon to signal the start of scanning. The system successfully detects the target (F1) when the confidence level of its bounding box exceeds the empirically set threshold of $0.3$. The system captures a key frame that includes both RGB and depth information. At the same time, another earcon sounds, signaling the user to pause and orient themselves toward the target object. This frame is used to calculate localization information (F2) and to query for further intent-based, long-text feedback. If the target object is not detected within a time limit of $45$ seconds, it is considered absent in the scenario. The user can choose to activate scene understanding (F4) or re-specify the target object.

\begin{figure}[t]
    \centering
    \includegraphics[width=1\linewidth]{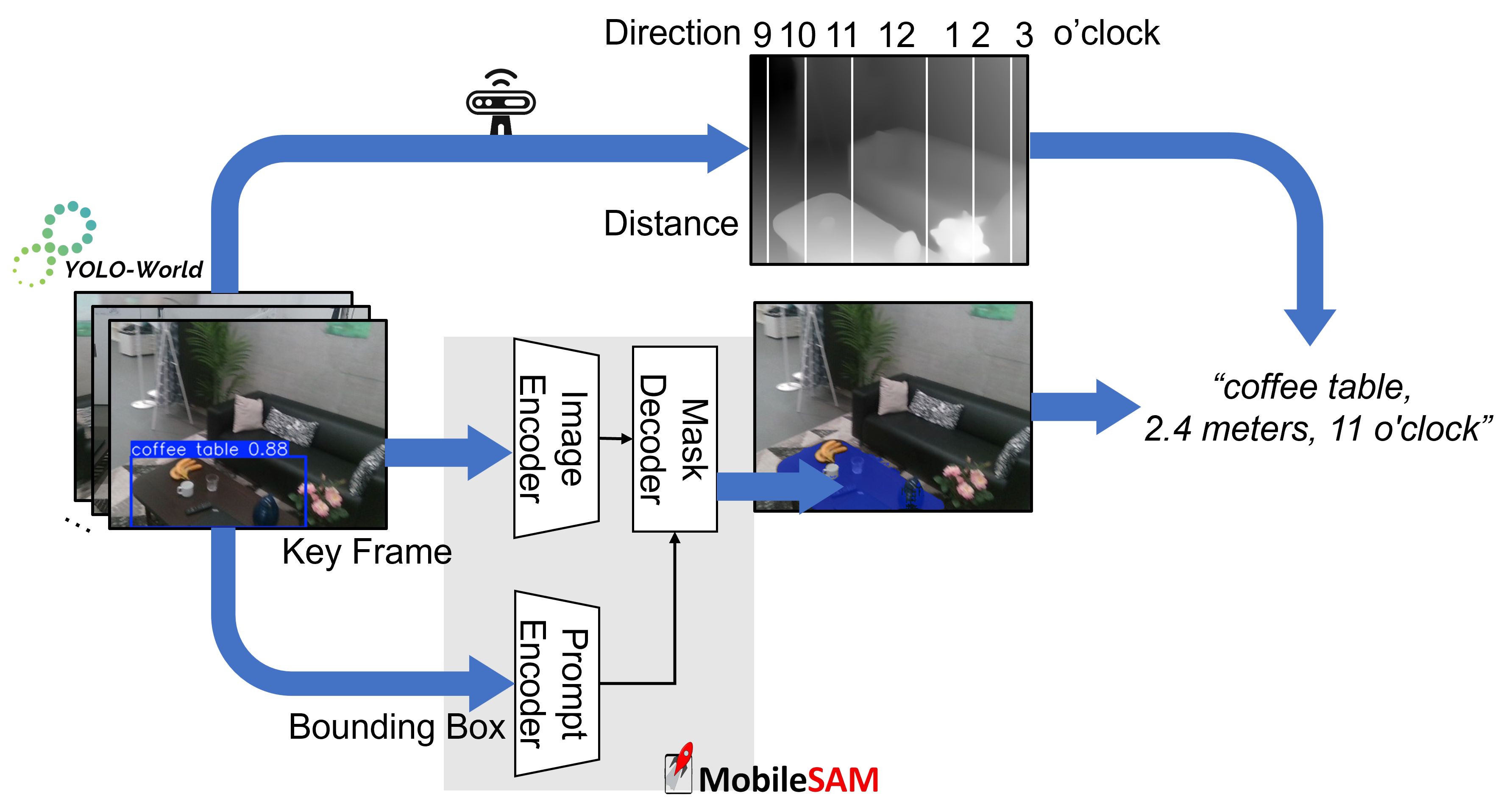}
    \caption{
    Object detection and localization: Each video frame is processed by YOLO-World to detect key frames in which the confidence level of the bounding box around the target object exceeds a certain threshold. Subsequently, the segmentation map generated by the bounding box, combined with the depth map, is used to provide precise localization information, including distance and direction.
    }
    \Description{The figure illustrates a pipeline for object detection and localization. A video stream is processed frame by frame using YOLO-World, which identifies objects and highlights keyframes where the target object is detected with high confidence (e.g., a coffee table with a confidence score of 0.88). The selected keyframe, including the bounding box, is passed into MobileSAM, a segmentation module. Within MobileSAM, the image and bounding box are processed through an image encoder and prompt encoder, respectively, then combined in a mask decoder to produce a segmentation map highlighting the target object. This segmented output is then fused with a depth map generated from the corresponding video frame to extract accurate localization information, such as the object's distance and direction from the camera.}
    \label{fig:localization}
\end{figure}

If the target is detected successfully, the egocentric localization information will be calculated using the keyframe and delivered in the format \textit{Object-Distance-Direction}, as proposed by Constantinescu~\textit{et al.}~\cite{Constantinescu2022listening}, as illustrated in Figure~\ref{fig:localization}. Egocentric information is presented in a clockwise orientation and distances in meters. 
Calculating distance using a bounding box is inaccurate.
The bounding box for object detection is considered accurate if it overlaps with the ground truth by at least $50\%$. However, this criterion might result in the bounding box inaccurately encompassing significant background areas~\cite{liu2023opensu}.
To enhance accuracy, MobileSAM~\cite{mobile_sam}, the compact version of SAM~\cite{kirillov2023segment}, was later added to generate segmentation masks $M$ using the bounding boxes as prompts post YOLO-World~\cite{Cheng2024YOLOWorld}. The distance of the object is calculated as the average depth from the key frame's depth map, masked by \( M \).
\begin{equation}
Distance = \frac{\sum (\text{depth\_map} \odot M)}{\sum M}
\end{equation}
To determine the clockwise direction, we use the center of the bottom edge of the frame, denoted as \( (x_c, y_c) \), as the clock's center. The angle \( \theta \) between the center of the bounding box \( (x_{\text{bbox}}, y_{\text{bbox}}) \) and the clock's center, relative to the bottom edge, is calculated as follows:
\begin{equation}
\theta = \arctan\left(\frac{y_c - y_{bbox}}{x_c - x_{bbox}}\right) \times \frac{180}{\pi}
\end{equation}
Then, \( \theta \) in the range \((-90^\circ, 90^\circ)\) is mapped to the clock positions from 9 to 3 o'clock.

\subsubsection{Module 3: System generates feedback.}

When generating feedback, the user is halted by an earcon and oriented towards the target object. Simultaneously, the keyframe capturing the user's egocentric view, which includes the target object, is sent to the MLLM to produce long-text feedback.

If the user considers the detected candidate to potentially be the target object, they may wish to learn more based on their intent. 
During the refinement iteration with the co-designer, we observed that he primarily had two intents regarding detected objects: \textit{navigating} to functional objects for interaction, such as finding a charger to charge a smartphone, and \textit{perceiving} regions of interest, such as identifying items on a tabletop without interacting with them. This was followed by a request for further details. Therefore, we have currently implemented branches for these intentions in this module, with the possibility of adding more in the future.

\paragraph{Generating feedback based on user intent.}
In the process of generating intent-based feedback, the system initially uses the keyframe for the first query, while subsequent queries are based on the current egocentric frame of the user. In the \textit{navigation} branch, the user initiates route planning (F3). After moving several steps, the target object may no longer be in the field of view, which could prevent further route planning. Consequently, the user has the option to either repeat the instruction or trigger a scene description (F4) for orientation. For example, if the user is aware that there is a desk on the route to the fan and understands from the scene description (F4) that the desk is directly in front of them, they will know \textit{``I'm getting close to the fan.''} If the user still feels lost, they can revert to the target specification to relocate the target object. In the \textit{perception} branch, the user can opt to use scene description (F4) to detail the surroundings of the target object, or directly ask open questions (F5) to engage in a conversation about the target object and its surroundings over several rounds. 
As the co-designer always discovered objects of interest or rejected the candidate after learning about the detailed information surrounding it, the user can respecify the target object at any step with ObjectFinder.

\paragraph{Optimizing interaction features.} 
Following the approaches of \cite{milios2018intelligenteye, Nourhan2020smart, marion2022wearable}, we have programmed the two buttons on the waist bag to select functions. We opted for this method over speech commands. Despite the flexibility, it is susceptible to environmental noise and the system's comprehension limitations~\cite{oumard2022voiceui}. Speech input is used only for target specification and dialogue in open questions (F5) for efficient object search.

\paragraph{Enhancing feedback accessibility through prompt engineering.}
According to VIALM~\cite{zhao2024vialm}, GPT-4 is the state-of-the-art for guiding blind people, excelling in both human (correctness, actionability, fluency) and automated evaluations. Thus, we chose GPT-4 for implementing MLLM functions (F3-F5). Effective prompt engineering is crucial for enhancing large language models' utility and accessibility for blind users. Our system prompt is concrete, incorporating \textit{role}, \textit{tone}, and \textit{response length}, guided by OpenAI’s Prompt Engineering tutorial~\cite{OpenAIPromptEngineering}. The response length, aligned with the alt text limits of social media (100-500 characters)~\cite{perkinsAltText}, is set to the maximum to accommodate user preferences for vivid responses. 
The co-designer scans the environment by turning his head rather than his body. When a target object is captured, he always pauses in his current posture, with his head and body often misaligned. Therefore, if the system provides egocentric instructions like \textit{``Please walk two steps forward.''} or \textit{``The desk is in front of you,''} it can lead to confusion.
To resolve this ambiguity, we precede each response from the MLLM with the instruction, \textit{``Please align your body with the direction of your head.''} This adjustment helps tailor the system's feedback to accommodate any user's scanning strategy. For route planning, we specifically consider that instructions based on \textit{steps} are easier for blind people to understand, as suggested by~\cite{bhanuka2023want}. Additionally, we provide instructions using landmarks rather than turn-by-turn directions, as per~\cite{jain2023want}. The system prompt and two user prompts for route planning (F3) and scene description (F4) are detailed in the supplementary material.

\begin{table*}
    \centering
    \caption{Demographics of participants. P0 is the co-designer who helped to adapt the system to the needs of the target group. P1-P8 were participants of the user study.}
    \setlength{\tabcolsep}{4mm}
    \resizebox{0.95\textwidth}{!}{
    \begin{tabular}{c|l|l|l|l}
    \toprule
         \textbf{User ID}& \textbf{Gender}& \textbf{Age Range}& \textbf{Vision Level, Onset}& \textbf{Experience of Apps}\\
    \midrule
         P0& Male&30-39&Light perception, since about 2004&BeMyAI, Seeing AI\\ \hline
         P1& Female&20-29&Light perception, since about 2022&BeMyAI, Seeing AI\\
         P2& Male&50-59&Fully blind, since birth&BeMyAI, Seeing AI\\
        P3& Male&20-29&Fully blind, since birth& BeMyAI\\
        P4& Male&20-29&Fully blind, since 2010&Lookout\\
        P5& Female&30-39&Light perception, since birth&Seeing AI, Envision\\
        P6& Female&50-59&Fully blind, since about 1989&Seeing AI\\
        P7& Male&70-79&Fully blind, since birth&None\\
        P8& Male&30-39&Light perception, since 2015&Seeing AI\\
    \bottomrule
    \end{tabular}
    }    
    \label{tab:participants}
\end{table*} 

\begin{figure*}
    \centering
    \includegraphics[width=\textwidth]{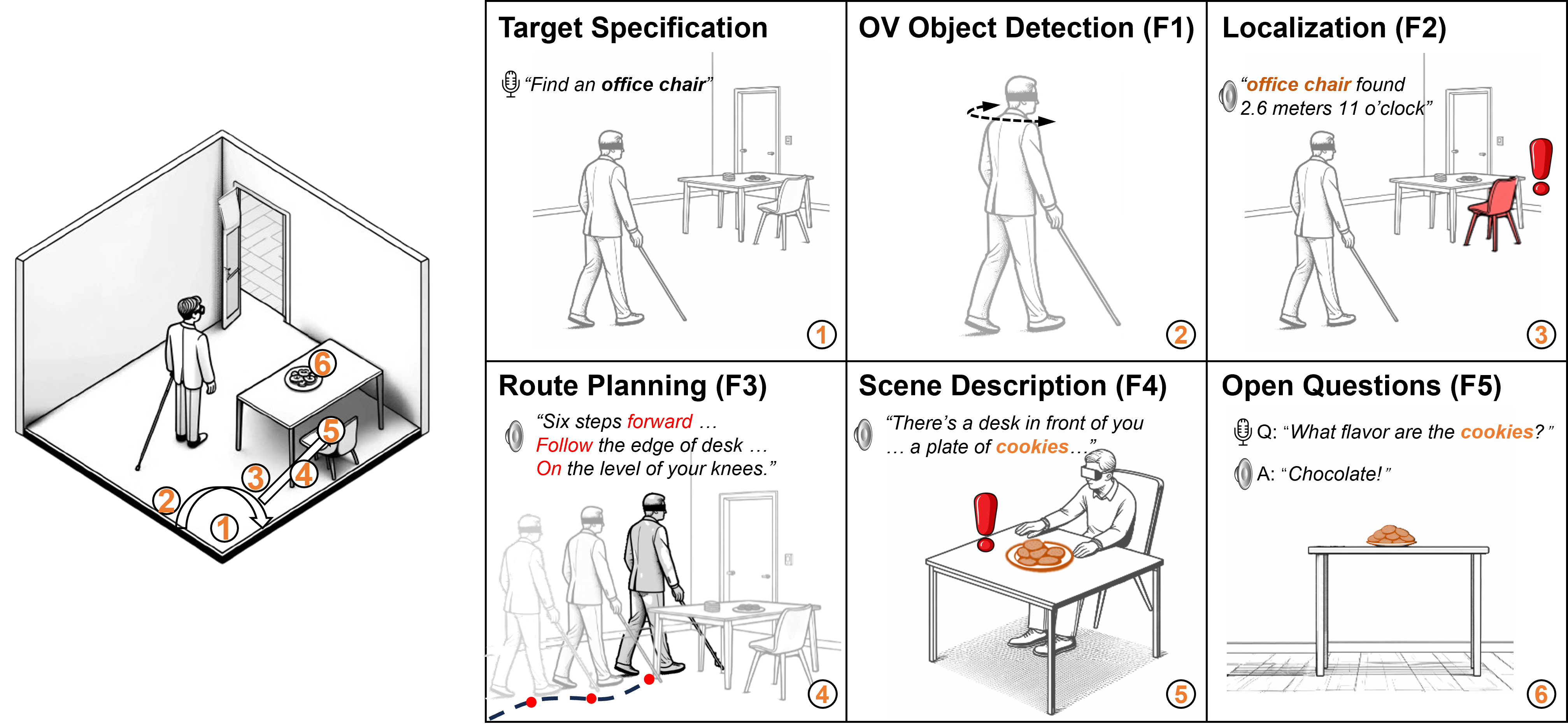}
    \caption{
    Simplified user study procedure focusing on two target objects: a large piece of furniture (an office chair) and a smaller object (a plate of cookies). The user walks into an unfamiliar environment, the office in this example. To sit in front of a desk, he or she should find an office chair first and navigate to it. 
    Then the user sits on the office chair and defines the desk as the region of interest. The user will know what is on the desk through the scene description function, a plate of cookies in this example, and get additional information through open questions.
    }
    \Description{This figure illustrates a bird's-eye view of a room layout and the sequential steps a blind user takes to navigate and interact with the environment using an assistive system. Step 1 involves the user entering the room and commanding the system to `Find office chair'. Step 2 initiates a scan of the environment for open-vocabulary object detection. In Step 3, the system provides localization information, indicating `office chair found, 2.6 meters, 11 o'clock.' Step 4 involves route planning with guidance like `Six steps forward, follow the edge of the desk, at knee level.' Step 5 allows the user to sit on the office chair and triggers a scene description: `There’s a desk in front of you with a plate of cookies.' Finally, Step 6 involves prompting the user to ask open questions such as `What flavor are the cookies?', to which the system responds `Chocolate!'}
    \label{fig:user_study}
\end{figure*}

\section{User Evaluation}
\label{sec:phase3}

In order to understand how ObjectFinder can support object search in unfamiliar environments, we conducted an exploratory study with eight blind people. In this step of our work, we include our own prototype ObjectFinder, ObjectFinder\_Base (a  
closed-vocabulary baseline prototype),  
and the commercial systems BeMyAI~\cite{BeMyAI2023} (a description-based application) and Lookout~\cite{lookout} (a detection-based application) as points of reference for participant feedback. Specifically, we focus on the following research questions:
\begin{enumerate}
    \item[RQ1:] To what extent does ObjectFinder deliver the necessary information for effective object search?
    \item[RQ2:] How do blind people perceive the
detailed scene context that ObjectFinder
provides to facilitate object searches?
    \item[RQ3:] How do blind individuals perceive ObjectFinder generally and in comparison to description- and detection-based systems?
    \item[RQ4:] What requirements do blind users consider important for an object search system, as their experiences with ObjectFinder show?
\end{enumerate}

\subsection{Participants and Procedure}\label{sec:5.1.1}
We recruited eight participants (P1-8 in Table~\ref{tab:participants}) from the local community using an existing mailing list. The participants ranged in age from $20$ to $80$ years ($\mu=40.75$ years, $\sigma=17.945$), including three women and five men. All participants were legally blind (vision$\leq 5\%$ for both eyes~\cite{WHO2021ICD10}), with seven having acuity $\leq 2\%$. Four of them were born blind.
For scene understanding, six of the participants had previous experience using description-based applications such as Seeing AI, BeMyAI, and Envision, while only one participant uses a detection-based application, Lookout. In Table~\ref{tab:participants}, we consider only the scene description or exploration feature of these applications, while usage of the applications for other purposes, such as reading documents, is not included.  Our study was approved by the university’s Ethics Committee. The video and audio recording were consented to by the participants.

Each user study lasted about two hours and consisted of the following steps: 
(1) an introduction and tutorial of our prototype; 
(2) exploration of both scenarios, office (\SI{7.95}{m^2}) and living room (\SI{15.96}{m^2}), using ObjectFinder and ObjectFinder\_Base interchangeably in a crossover manner~\cite{jones2014crossover}, 
each followed by (3) the completion of a questionnaire featuring Likert-scale evaluations of function and the NASA-TLX~\cite{hart2006nasa} for assessing cognitive load, followed by a short semi-structured interview; (4) short exploration of the living room scenario using the commercial applications BeMyAI and Lookout, followed by (5) another short semi-structured interview. 

\subsection{Scenario Exploration} 
\label{subsec:scene_exploration}
\begin{figure*}[htbp]
    \centering
    \begin{subfigure}[b]{0.49\textwidth}
        \centering
        \includegraphics[width=\textwidth]{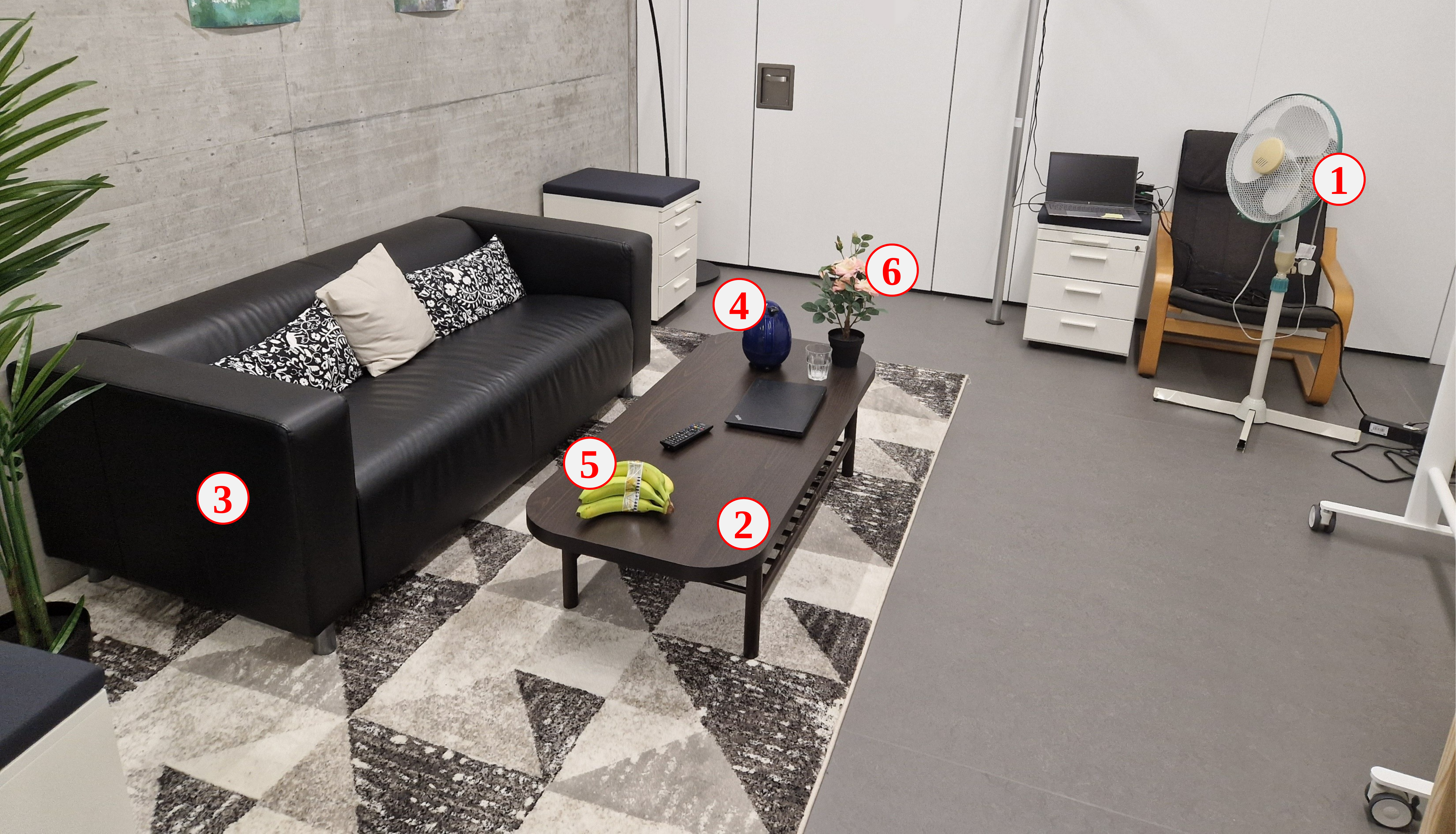}
        \caption{Living Room.}
        \label{fig:living_room}
    \end{subfigure}
    \hfill
    \begin{subfigure}[b]{0.49\textwidth}
        \centering
        \includegraphics[width=\textwidth]{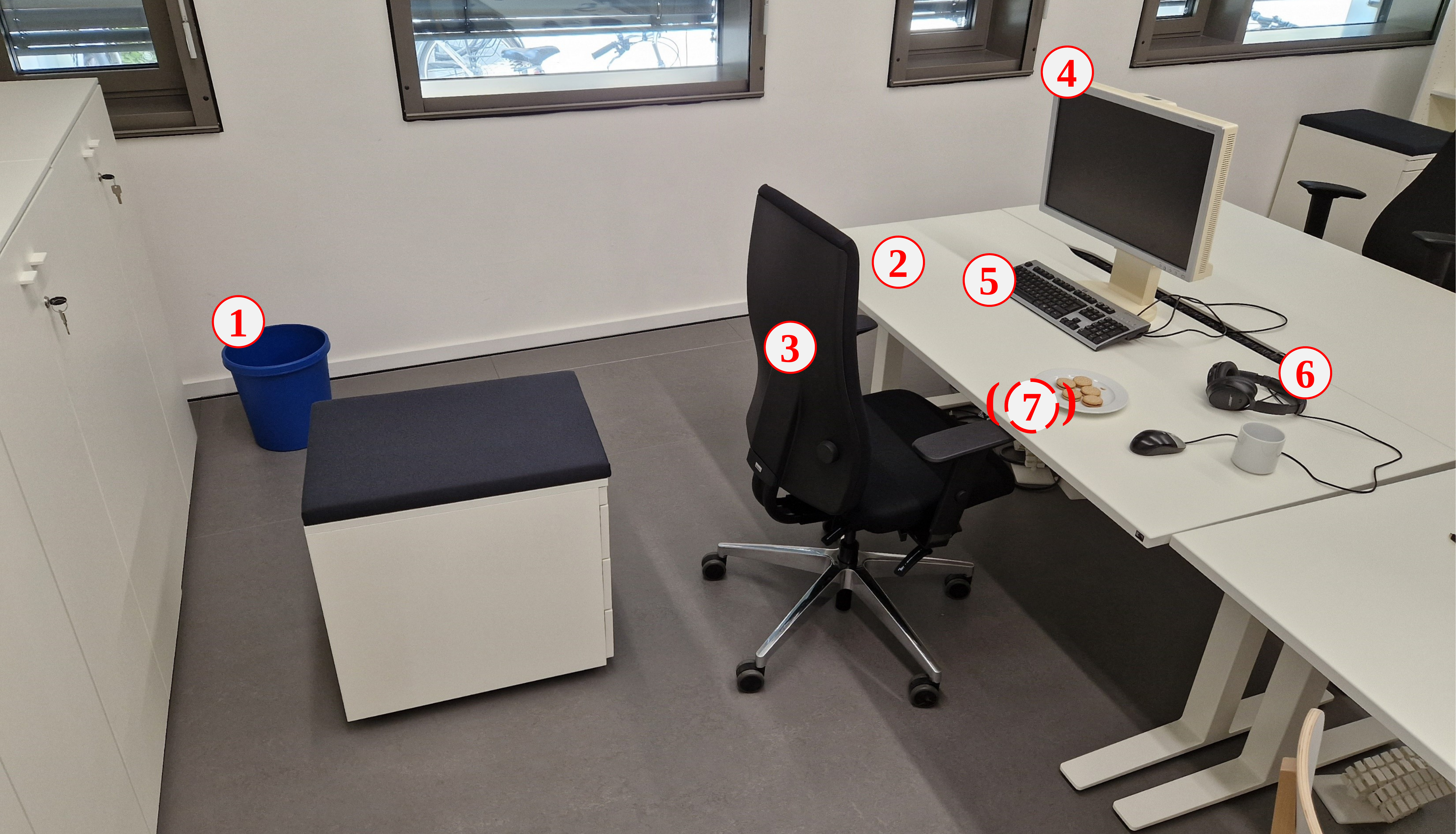}
        \caption{Office.}
        \label{fig:office}
    \end{subfigure}
    \caption{Layout of the two scenarios and the order of target objects. The 7th object in the office, \textit{a plate of cookies}, is the bonus, to determine if participants are aware of its existence through the system. The photos are taken at the starting points of the task in each scenario.}
    \Description{This figure provides a visual representation of the layout for two scenarios: a living room and an office. It details the sequence and locations of various target objects within these settings.}
    \label{fig:scenarios}
\end{figure*}

\begin{figure*}
    \centering
    \includegraphics[width=0.9\textwidth]{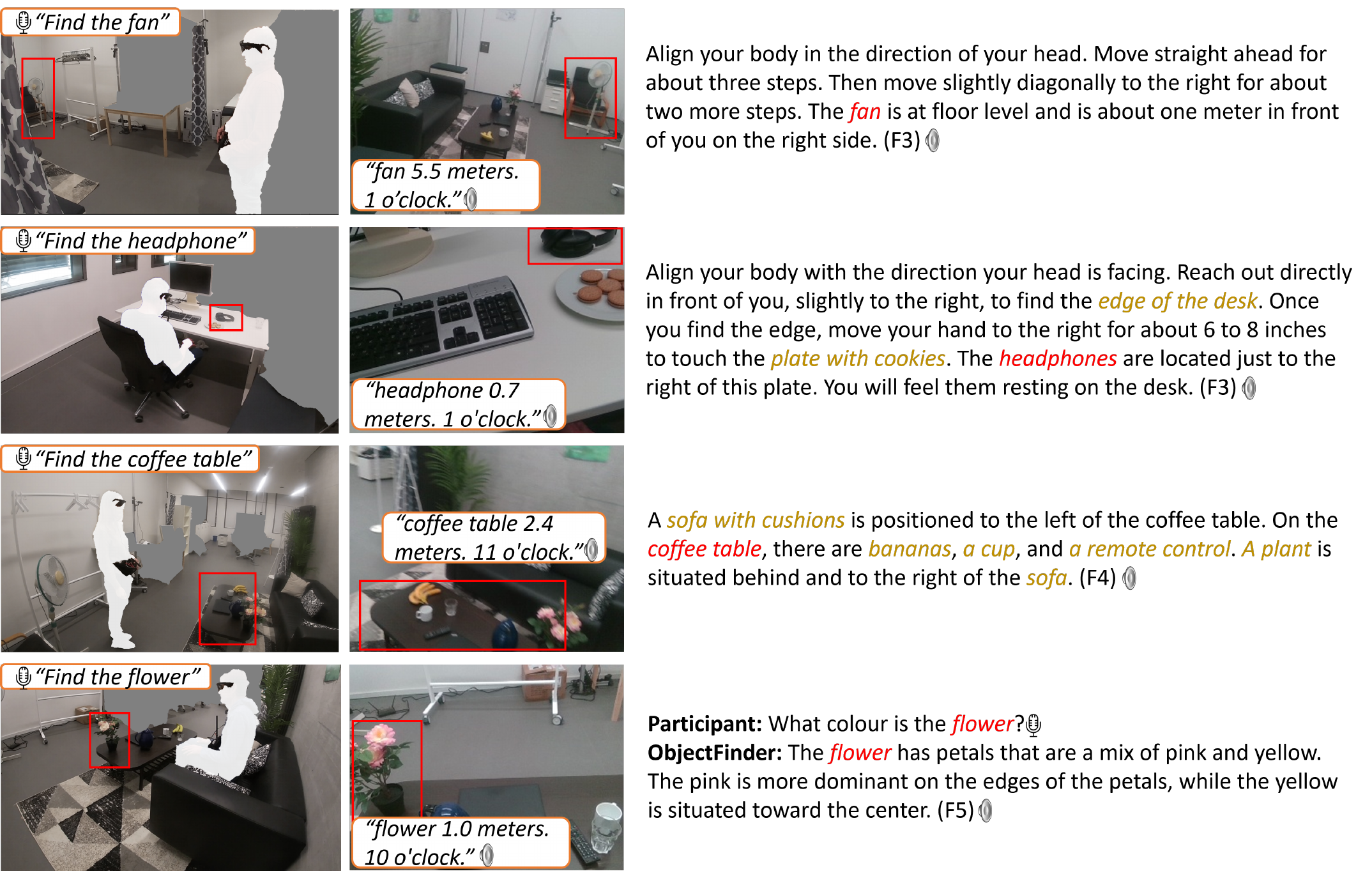}
    \caption{
The first column shows examples of postures, while the second column displays key frames captured when target objects were detected (F1) along with the corresponding real-time egocentric localization information (F2). The third column presents system feedback generated by the MLLM, which uses route planning (F3) to reach both large and small items, employs scene description (F4) to describe the coffee table, and utilizes open questions (F5) to gather additional details.
    }
    \Description{This figure presents four real examples from a user study illustrating how participants navigate to and interact with both large pieces of furniture and small objects, as well as how they describe and inquire about items in a room setting.
    Example 1 involves navigating to a fan (5.5 meters, 11 o'clock) with detailed route planning instructions: 'Align your body in the direction of your head. Move straight ahead for about three steps. Then move slightly diagonally to the right for about two more steps. Then fan is at floor level and is about one meter in front of you on the right side.'
    Example 2 describes finding headphones detected at (0.7 meters, 1 o'clock) with instructions: 'Align your body with the direction your head is facing. Reach out directly in front of you, slightly to the right, to find the edge of the desk. Once you find the edge, move your hand to the right for about 6 to 8 inches to touch the plate with cookies. The headphones are located just to the right of this plate. You will feel them resting on the desk.'
    Example 3 details the items on and around a coffee table (2.4 meters, 11 o'clock): 'A sofa with cushions is positioned to the left of the coffee table. On the coffee table, there are bananas, a cup, and a remote control. A plant is situated behind and to the right of the sofa.'
    Example 4 shows an inquiry about a flower (1.0 meters, 10 o'clock): Participants ask 'What colour is the flower?' and ObjectFinder responds 'The flower has petals that are a mix of pink and yellow. The pink is more dominant on the edges of the petals, while the yellow is situated toward the center.'}
    \label{fig:key_frames}
\end{figure*}

Blind people typically search for large items as landmarks to construct mental maps of unfamiliar environments, and they would like to use the system to explore small objects on the tabletop.
In each scenario (living room or office), participants were asked to find six target objects: 
three large pieces of furniture to establish spatial understanding, followed by three smaller objects found on the coffee table or desk. 
Figure~\ref{fig:user_study} describes a simplified procedure involving the five functions F1-F5. 
Table~\ref{tab:target_objects} specifies the target objects that need to be found in sequence and the initial MLLM functions to be triggered when these target objects are detected.
The layout and the order of targets for search are shown in Figure~\ref{fig:scenarios}. 
We are the first to engineer prompts that generate route planning instructions for blind individuals, guiding them to objects. Therefore, our primary focus is on testing the route planning function (F3).
Since the closed-vocabulary ObjectFinder\_Base can only detect a limited number of target objects~\cite{lin2014microsoft}, we categorize the six target objects in each scenario into three groups: \textit{unrelated to} (two objects), \textit{related to} (three objects), and \textit{exact in} (one object) COCO2017. As for the \textit{related} targets, we will hint to the participants to look for the related objects in COCO2017 when they are using the baseline, but they will experience a vocabulary gap between the target specification and MLLM feedback. As for the \textit{unrelated} objects, the participants cannot even specify the target objects \textbf{(C1)}. So we provide the option for the participants to trigger the scene description function several times to find the objects.
\textit{Cookies} are the bonus target, and we observe whether the participants can recognize it themselves to validate the capability of our system to inspire the detection of unexpected targets \textbf{(C2)}. Figure~\ref{fig:key_frames} illustrates examples of how the participants detected the target objects and received the system feedback during the user study. 

Lookout and BeMyAI, commercial applications familiar to participants but not designed for object search, were briefly used in the living room scenario to locate a \textit{fan} and a \textit{teapot} as reminders of their functions.

\begin{table*}[ht]
    \centering
    \caption{Target objects in each scenario. The superscript \texttt{object$^{F}$} denotes the function that is activated first upon finding the target objects. F3: route planning; F4: scene description; F5: open questions.}
    \begin{tabular}{l|l|l|l}
    \toprule
         \textbf{Scenarios} & \multicolumn{2}{c|}{\textbf{Target Objects}} & \textbf{Names in COCO~\cite{lin2014microsoft} dataset} \\
    \hline
         \multirow{2}{*}{Office} & Furnitures & trash bin$^{F3}$, desk$^{F4}$, office chair$^{F3}$ & <None>, dining table, chair \\
    \cline{2-4}
         & Smaller objects & monitor$^{F5}$, keyboard$^{F3}$, headphone$^{F3}$ (cookies$^{F3}$) & TV, keyboard, <None> (<None>) \\
    \hline
         \multirow{2}{*}{Living Room} & Furnitures & fan$^{F3}$, coffee table$^{F4}$, sofa$^{F3}$ & <None>, dining table, couch \\
    \cline{2-4}
         & Smaller objects & teapot$^{F3}$, banana$^{F3}$, flower$^{F5}$ & <None>, banana, potted plant \\
    \bottomrule
    \end{tabular}
    \label{tab:target_objects}
\end{table*}

\subsection{Data Analysis}
We have both qualitative and quantitative data. For qualitative data, the user study transcripts were analyzed using the hybrid process of inductive and deductive thematic analysis proposed by Fereday and Muir-Cochrane~\cite{FeredayMuirCochrane2006}.
The first author led the analysis by repeatedly reading the transcript for familiarization and coding it in multiple rounds.
Beyond data-driven inductive coding, we also applied deductive coding, which yielded meaningful insights into the system’s capacity to identify regions of interest \textbf{(C1)} and to facilitate the discovery of unexpected targets \textbf{(C2)}.
In a workshop, the research team assigned $243$ data points to $69$ codes, which were further refined to $12$ codes, and finally, four themes were crafted and will be presented in Sec.~\ref{sec:5.2}. For quantitative data,  
we limit our comparison to descriptive statistics due to the small size of the user group.

\section{Findings}\label{sec:5.2}
\subsection{RQ1: To what extent does ObjectFinder deliver the necessary information for effective object search?}
\textit{Participants found that ObjectFinder provides an adequate amount of information, including the obligatory egocentric (distance and direction) and allocentric (relationships among the objects) information, for object search.  On the other hand, participants have varying perceptions of the optional information (\eg~ color and alert information).
}

\paragraph{Amount of Information}
Regarding the amount of information provided, three participants found it adequate for their needs. The average system feedback for route planning (F3) and scene description (F4) contains $62.90$ words, with a standard deviation of $19.11$. System feedback for open questions (F5) is relatively shorter. However, when asked about their preference for more or less information, responses were divided: half preferred more, while the other half favored less. We noted that preferences for information quantity relate to individual processing styles: while some participants could ignore excess information, others felt overwhelmed. The participants chose to receive more information, generally expressing a preference for as much as possible. As one participant noted, \textit{``as much as you can get. It's completely blank for me, so the more I have, the better.''} (P5). P4 expressed that \textit{``things that don't interest me, I can just ignore.''}
Conversely, P6 and P7 explained their preference for less information, attributing it to not being accustomed to processing such a large amount of visual information and feeling overwhelmed by it. 
P8 suggested that the information provided could perhaps be reduced after getting the first overview: \textit{``it could maybe be a little bit less, so if you know once there are some objects on the table, then you don't really need this information the second or third time unless you ask the system what's on the table.''} 
Additionally, we observed that the amount of information varies across functions. BeMyAI, used solely for scene description, received praise from three participants for its \textit{``detailed''} information, although one deemed it excessive. In contrast, participants noted that our ObjectFinder delivered more information than necessary for efficient navigation.

\paragraph{Obligatory Information}
Participants (P6-8) highlighted the importance of localization information (distance and direction) after scene exploration.
The interpretation of the distance should be clear and intuitive. 
P1 highlighted the usefulness of different distance measures 
noting, \textit{``I think the steps are good for like when it's really near, so it's just a few steps. But if there are longer distances. I think sometimes meters might be more useful [...] 
So it sometimes has like $0.1$ meter [...] you could have said `right in front of you' or `one step' like that.''}
Participants mentioned that the distances reported were inaccurate, appearing larger than they perceived. Upon reviewing the video, we observed that some discrepancies were caused by measurements taken from the head to the object, rather than horizontally.
This inaccuracy may be more pronounced in our small indoor settings.
To describe the direction, P5 suggested to \textit{``specify a $20$ centimeters margin''}, noting that if an object is $10$ centimeters to the left or right, it's still considered in the front. 

Contextual information around the target is crucial for object search, such as relationships among surrounding objects and their location information. For example, P3 noted, 
\textit{``it's good that the objects around it are announced, so you have some idea where the object you want to find is in relation to other objects.''}
A description should also include the distance, which says something about the user as a reference point to the object. P8 suggested to improve ObjectFinder by incorporating distance into the MLLM output,
\textit{`` when you make the description, it holds everything, but it does not really talk about the distance.''}

\paragraph{Optional Information}
Participants expressed mixed feelings about the relevance of color and alert information, which emphasizes the importance of personalization. 
For example, P5 and P6 appreciated the unsolicited color details. \textit{``It told me the color of the remote control without my asking [...] it helps me to visualize the environment around.''} (P5). In contrast, P2 criticized the excessive information, remarking, \textit{``I want to walk from A to B, and I don't want a literature presentation of the color and the landscape description which can fill books.''}
Regarding the alert information, P2 valued the clarity of certain descriptions: \textit{``what I liked was the description. It directly tells people to be careful `move your arms', and I think it was a very clear description.''} However, P3 found it superfluous, commenting, \textit{``about the whole flavor text about `not bumping into an object', `sitting down', `carefully turning the chair toward me first'. This is all not really necessary.''} 
\begin{figure*}
    \centering
    \includegraphics[width=0.9\textwidth]{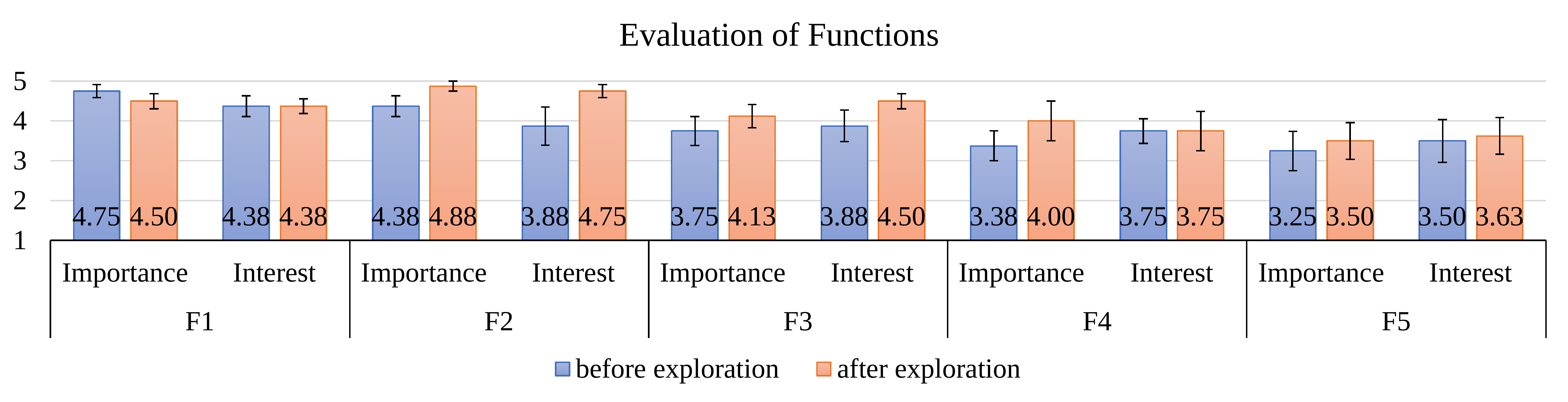}
    \caption{Evaluation of the five functions in terms of importance and level of interest, using Likert-scale scores, before and after exploring two scenarios. F1: object detection; F2: localization information; F3: route planning; F4: scene description; F5: open questions. }
    \Description{This bar chart compares Likert-scale scores for five functions before and after an exploratory user study, assessing importance and level of interest. Most scores increased post-study, except for object detection, which slightly decreased from 4.75 to 4.5. Localization information received the highest scores, with importance rising from 4.38 to 4.88 and interest from 3.88 to 4.75. Scores for the first three functions exceeded 4, while the remaining two stayed below 4.}
    \label{fig:functions}
\end{figure*}

\subsection{RQ2: How do blind people perceive the detailed scene context that ObjectFinder provides to facilitate object search?}

\textit{Participants were able to gain an overview of the scene, orient themselves, and explore search options within the detailed scene context provided by ObjectFinder.}

There is some evidence that participants obtained more detailed environmental information through the use of ObjectFinder.
P1 noted:
\textit{``I think it was very good [to know] where I am. When it said where my target object was, it also told me the surrounding objects, because maybe if I find the surrounding object first, then I know, OK, I'm close.
''}
P6 noted the benefits of discovering other usable items \textbf{(C2)} and understanding their arrangement,
\textit{``I didn’t know there was another chair there. It gives more information about objects and their arrangement.''} Participants would like to get an overview of unfamiliar rooms without receiving too many details. P8 explained, \textit{``I could imagine if you really don't know the room, and you want to first [have an] overview of what is in the room, it was pretty detailed.
''}

ObjectFinder's ability to locate regions of interest is a key feature that facilitates detailed descriptions, aiding in navigation and discovery without physical search \textbf{(C1, C2)}. 
P6 particularly valued this feature, noting:
\textit{``if it's a new desk, I don't have to feel around to know where the computer is. I don't have to move far right or left to find the position easily.''} 
Similarly, P1 found the system's detailed output helpful for discovery \textbf{(C2)}, 
\textit{``When I first got into the scene, I got a very detailed description of the coffee table [...] so I knew what I could expect to find there.''} Moreover, the wearable design of ObjectFinder with glasses that capture egocentric images significantly improves participants' sense of direction \textbf{(C1)}.  \textit{``I found it easier to think about which direction.''} (P6). 

\subsection{RQ3: How do blind individuals perceive ObjectFinder generally and in comparison to description- and detection-based systems?}

\textit{Participants generally prefer ObjectFinder for object search over description- and detection-based systems because it enables users to specify targets actively and provides both egocentric and allocentric essential information, though some practice is required to become accustomed to it.}

\paragraph{General Perception of ObjectFinder}
Generally, five out of eight participants expressed excitement about ObjectFinder after exploring it. \textit{``I was very positively surprised at how good it works, and how easy it is to get information and to find myself in the scene.''} (P1). 
ObjectFinder was also appreciated for accurate object searches without physical touch. 
Subsequently, participants rated the system's helpfulness ($\mu=4.13, \sigma=0.641$) and their independence ($\mu=4.31, \sigma=1.01$) with ObjectFinder in an unfamiliar environment, with both ratings based on $5$-point Likert scales and averaged over $4$.
Through ObjectFinder, participants discovered more than just target items, finding inspiration in unexpected objects, both small and large \textbf{(C2)}. 
For example, \textit{``I'm sort of inspired because now I know what options I have, instead of walking in and just looking for my phone because I know it must be there, but I don't know what else could be there [...] I was inspired to look for, for example, bananas and stuff like that.''}. (P1).
However, P3 noted that searching with ObjectFinder might be slower than tactile exploration in our compact indoor setting, noted, \textit{``there are so many steps putting it up and searching for the thing, and maybe it doesn't even find it (ObjectFinder\_Base).''} The search could be accelerated by using ObjectFinder alongside the cane, taking advantage of the cane’s large radius (P2).

\paragraph{User Preferences and Cognitive Load}
Regarding the differences between the commercial applications and the prototypes, 
participants valued that they could actively define the target objects with ObjectFinder, making it more reliable in searching for specific items. As P6 mentioned, \textit{``what I also liked about this system is speaking to it, I find that more targeted.''}. P3 echoed this sentiment, highlighting the biggest advantage: \textit{``The biggest advantage [...] contrary to BeMyAI is that you can specify what you want to find.''} \textbf{(C1)}. 
"Furthermore, as previously mentioned, ObjectFinder delivers essential information for object searches that cannot be provided by either description-based or detection-based systems.
This was more difficult using the existing commercial systems. For example, after using BeMyAI, P8 mentioned, \textit{"BeMyAI had no information regarding the distances, so you have 
[...] a less sense of location
"}. As previously analyzed, egocentric localization information (distance and direction) and the allocentric relationships among surrounding objects are crucial for effective object search. The lack of localization information (P2, P8) and the absence of context regarding object relationships (P2) were mentioned as reasons why participants did not prefer Lookout, highlighting the potential for solutions that can combine multiple aspects of object search.

As shown in Table~\ref{tab:nasa_tlx}, the study also examined cognitive load, showing that it was relatively low across system types and scenarios. 
Retrieving keyframes for MLLM information and participant feedback reveals that the landscape orientation of the camera on the glasses, contrasting with participants' usual practice of taking portrait photos using commercial applications, and the presence of relatively low furniture in the living room often result in the prototype failing to recognize the \textit{coffee table} as an obstacle to the \textit{sofa}.
The exploratory experience with the prototype was not comprehensive enough.
P7 mentioned that F4 and F5 are insufficiently tested in our study, and added: \textit{``one needs to get used to it, it requires some practice, and then the search will be very precise.''} P5 mentioned that she was not accustomed to the earcon indicating to stand still.

\begin{table*}[ht]
    \centering
    \caption{Assessment of workload required to complete tasks with two systems (ObjectFinder \textit{vs.} ObjectFinder\_Base) and in two scenarios (Office \textit{vs.} Living Room) using NASA-TLX. The scores range from 1 (very low) to 21 (very high).}
    \begin{tabularx}{\textwidth}{c|>{\centering\arraybackslash}X|>{\centering\arraybackslash}X|>{\centering\arraybackslash}X|>{\centering\arraybackslash}X|>{\centering\arraybackslash}X|>{\centering\arraybackslash}X|>{\centering\arraybackslash}X|>{\centering\arraybackslash}X}
    \toprule
         \multirow{3}{*}{\textbf{Sub-scale}}&\multicolumn{4}{c|}{\textbf{Systems}}&\multicolumn{4}{c}{\textbf{Scenarios}}\\

         &\multicolumn{2}{c|}{ObjectFinder}&\multicolumn{2}{c|}{ObjectFinder\_Base}&\multicolumn{2}{c|}{Office}&\multicolumn{2}{c}{Living Room}\\

         &Mean&Std Dev&Mean&Std Dev&Mean&Std Dev&Mean&Std Dev  \\
         \midrule

         Mental Demand&4.25&2.96&\textbf{5.38}&3.42&4.50&2.93&\textbf{5.13}&3.52\\
         Physical Demand&2.44&1.76&\textbf{3.00}&2.39&2.69&2.25&\textbf{2.75}&1.98\\
         Temporal&\textbf{2.63}&2.00&\textbf{2.63}&1.30&2.13&1.46&\textbf{3.13}&1.73\\
         Performance&4.69&2.02&\textbf{7.25}&2.38&4.94&2.43&\textbf{7.00}&2.27\\
         Effort&4.13&2.03&\textbf{5.38}&3.11&\textbf{4.75}&3.11&\textbf{4.75}&2.25\\
         Frustration&4.13&3.48&\textbf{4.38}&3.07&4.00&3.46&\textbf{4.50}&3.07\\
         
    \bottomrule
    \end{tabularx}
    \label{tab:nasa_tlx}
\end{table*}

\subsection{RQ4: What requirements do blind users consider important for an object search system, as their experiences with ObjectFinder show?} 

\textit{Participants basically suggested that efficiency is crucial in terms of interaction features and hardware for object search. Additionally, they considered object detection, localization, and route planning to be important functions of object search.}
\paragraph{Interaction Features.}

It seems that interaction features are assessed differently among participants.
P4 felt pressing buttons was faster than using a touchscreen, though half of the participants found the button codes for option selection unusual. 
Suggestions included maintaining a unified button combination for each function (P3) or using \textit{``four different pressings''} (P8). P1 preferred pressing the button without waiting for the entire list to be read aloud.

Regarding target specification, P8 described the voice commands as \textit{``pretty, pretty easy''}, while P4 expressed concerns with programs that solely accept voice input, suggesting \textit{``maybe it's better to have the ability to switch between a list of maybe recognized objects and voice commands.''} Besides the current earcons, P6 suggested creating an additional earcon for option confirmation, rather than using spoken text.

As mentioned before, some participants preferred more information, while they suggested \textit{``implementing a skip option''} (P1) and the ability to \textit{``switch information on and off optionally''} (P2). P4 suggested adding a main menu to easily switch between object search (detection and navigation) and scene description. 
The interaction feature of Lookout was highlighted by P1: \textit{``I don’t have to take a picture and wait for a response.''}

\paragraph{Software Requirements}
We asked participants to rate the functions before and after using ObjectFinder and ObjectFinder\_Base in the explored scenarios, in terms of their perceived importance and level of interest, as captured by Likert scores (Figure~\ref{fig:functions}). From the ratings, the functions object detection, localization information, and route planning are important for the object search task ($\mu{>}4.0$), 
accompanied by a reduced standard error. Among the five functions evaluated, localization information was rated as the most important. 

Compared with the commercial applications, P8 mentioned the advantage of ObjectFinder is \textit{``the distance and also the guiding function which is not really available for the other both apps.''} (P8). We note that the software should be fast and make fewer mistakes. P2 mentioned that receiving explanations from a sighted person on BeMyEyes \textit{``works better because it’s without delay.''} Half of the participants mentioned that frequent misidentifications of Lookout were not preferred. As P5 noted, \textit{``it misidentified objects [...], like there’s no dishwasher,''} and at times, the \textit{teapot} was categorized as a \textit{helmet} or \textit{mouse}. 
 
\paragraph{Hardware Requirements}

Participants noted the distinct advantages and disadvantages of capturing scenarios with glasses compared to a cellphone. Three participants experienced reduced mental effort in determining the camera's orientation when using glasses that capture the egocentric view. \textit{``It's always harder for me to think exactly about what the phone is capturing with the phone camera, I found that better with the glasses.''} (P6). 
On the contrary, users required external cues to detect objects in the lower part of their viewing field. 
\textit{``The possibility of overlooking some obstacles like the coffee table [...] if I hadn't known that there was the table, I would have just run into it, which is not that nice.''} (P1). To detect the obstacles, such as \textit{coffee table}, participants should lower their heads rather than step backward. 
Two participants, who explored the office using ObjectFinder\_Base, which couldn't locate the region of interest for certain objects, without lowering their heads, failed to detect items on the desk and remained unaware of a plate of \textit{cookies} also present there. \textbf{(C2)}
Additionally, using glasses poses a challenge since they represent an additional item to carry alongside smartphones. 
Consequently, a pair of glasses is another item for them to carry and potentially forget, alongside their smartphones (P4, P5). 
The price of the glasses was also mentioned to be considered.

\section{Discussion}
Our research highlights that object-search systems have potential as an assistive technology for people who are blind. Here, we want to discuss future directions for the development of such systems, outlining challenges and opportunity for the Human-Computer Interaction and accessibility research communities.

\subsection{Potential Features for Future Integration}
In our study, we uncovered a range of features and characteristics of ObjectFinder that offer interesting avenues for future work, either through iteration on our system, or integration in other, comparable systems. Here, we give an overview of the most relevant aspects.

\textbf{Providing better and more tailored descriptions.} Providing descriptive overviews of unfamiliar environments is essential for revealing unexpected objects and should be relative to the user's position. 
Based on our findings, we recommend to include both distance and direction in descriptions, with far objects quantified in meters and nearer objects described in steps or as \textit{``right in front of you''}. These details can facilitate room exploration, help in building mental maps, and assist in orientation.

\textbf{Understanding advantages and drawbacks of additional information.} Although the essential information for object search, localization information, and relationships among surrounding objects, is well-defined, participants had mixed feelings about additional details like color and alert information. Consequently, the future object-search system should allow users to select the amount of information 
by incorporating options to skip information or to stop and continue information output. Likewise, future work should explore user preferences for the types of information to be included, and implications of additional unrequested descriptions for user experience and aspects such as cognitive load.

\textbf{Improving system reliability and integration with other assistive technology.} Unsurprisingly, participants expressed a preference for a system that is error-free and can accurately locate searched objects. From a system requirements perspective, the camera should capture both nearby objects at a lower angle and objects at a distance. The system should be lightweight and offer intuitive interaction features. For our prototype, participants appreciated the use of a button for option selection over a touchscreen, as well as voice commands. Additionally, earcons should be intuitive, while a training session to acquaint users with the meanings of earcons at the outset is advisable. The design of the system should serve as an extension to the cane, taking advantage of its large radius.

\textbf{Addressing portability and social accessibility of the system.} Our results show that participants reflected on the hardware included in the current iteration of the prototype, which is clearly visible, and takes up significant space when attached to the user's body. Here, participants expressed a preference for a smartphone-based solution. On the one hand, this is a sustainable approach that leverages hardware already in the possession of users. On the other hand, this may address concerns with respect to social accessibility, \textit{i.e.}, the visibility of assistive technology to others, and associated stigma~\cite{kristen2016self}.

\subsection{Tensions and Concerns Regarding Vision- and AI-based Assistive Technology}

There are tensions and concerns that need to be resolved for systems such as ObjectFinder to effectively and safely support object searching.

\textbf{User habits and needs regarding lighting conditions for camera-based systems.} With respect to the technical requirements of current approaches to computer vision systems, we note that lighting conditions are crucial for camera-based systems. However, this conflicts with the fact that light plays a different role in the lives of blind people: Legally blind persons may not use light in their homes and workplaces in the same way as (typically) sighted system developers would anticipate ~\cite{Penuela2024usecase}, and people who do have residual vision may not find lighting conditions required by camera systems comfortable.

\textbf{Addressing safety concerns in the context of AI.} Likewise, we noted instances in which ObjectFinder did not recognize furniture, e.g., when living room furnishings were lower than those in office environments, and not in view of the camera worn by participants. While there are established strategies for people who are blind or have low vision to use gaze for environmental scanning~\cite{visionaware_scanning_2023}, no specific scanning strategies for smart glasses tailored to blind individuals currently exist. With advancements in wearable systems and smart glasses, researching and integrating these specific scanning techniques into mobility training is both viable and beneficial. Moreover, incorporating sensors and cameras with wider angles, like LiDAR~\cite{liu2021hida} and omnidirectional camera~\cite{kawaharazuka2024reflex}, could enhance scene perception over larger areas, reducing the need for physical scanning efforts. However, despite potential technological solutions (which may come with new challenges), this specific instance highlights tensions around safety: On the one hand, users are invited to rely on systems for object detection, on the other hand, it is known that vision- and AI-based systems can be unreliable in specific situations (e.g., in the context of autonomous driving~\cite{cummings2024unreliable}), and even more so in the context of disability (e.g., see~\cite{kane2020sense}). Thus, there remains a tension between what AI-based assistive technology seeks to offer, and what it can realistically provide, which is an aspect that needs to be communicated with nuance, and should be negotiated together with target audiences in the context of future work. 

\textbf{Understanding the limitations of technology for object search.} Finally, the Human-Computer Interaction and accessibility communities have previously discussed issues surrounding technologies focusing on the independence of users. In particular, Vincenzi~\cite{beatrice2021guiding} contributed a critical appraisal of assistive technology for navigation of blind people, suggesting that there were instances where working with other people was more relevant than applying a technology-based solution. In this context, the principle of interdependence~\cite{bennett2018independence} is relevant, i.e., the fact that we all exist within relationship with our environment, and that assistive technology should not only consider the individual user but also the (social) context within which it exists, and implications of its design under consideration of opportunities to create collective access. Here, we need to ask critical questions around specific features of ObjectFinder, and we want to leave you with one example: \textit{Is it really necessary for a blind person to use the system in the workplace, or could non-disabled colleagues make a bigger effort to not misplace or alter their desk?}

\section{Conclusion}
In this work, we explored the design and development of a prototype that combines detection and description to enable open-vocabulary interactive object search for blind people. With our prototype, we address shortcomings of existing systems that are either description- or detection-based: locating regions of interest and discovering incidental targets. The system feedback is tailored to various user intents, and our exploratory user study suggests that this approach is promising, as it provides essential egocentric localization and allocentric scene context while enabling interactive object search.
Overall, our work represents an initial step towards developing AI-based assistive technology that supports object search, providing first insights into user requirements and application challenges. 
Here, we hope that our work will encourage and facilitate further development of object-search systems, and that it will inspire future studies into the experiences that blind people have with such technologies.

\section{Acknowledgements}
This work was supported in part by the Ministry of Science, Research and the Arts of Baden-Wurttemberg (MWK) through the Cooperative Graduate School Accessibility through AI-based Assistive Technology (KATE) under Grant BW6-03, in part by funding from the pilot program Core-Informatics of the Helmholtz Association (HGF), and in part by Karlsruhe House of Young Scientists (KHYS). 
We thank HoreKA@KIT, HAICORE@KIT, and bwHPC supercomputer partitions. We adjusted in some cases the grammar of the paper using ChatGPT.

\bibliographystyle{ACM-Reference-Format}
\bibliography{main}


\begin{thebibliography}{100}


\ifx \showCODEN    \undefined \def \showCODEN     #1{\unskip}     \fi
\ifx \showDOI      \undefined \def \showDOI       #1{#1}\fi
\ifx \showISBNx    \undefined \def \showISBNx     #1{\unskip}     \fi
\ifx \showISBNxiii \undefined \def \showISBNxiii  #1{\unskip}     \fi
\ifx \showISSN     \undefined \def \showISSN      #1{\unskip}     \fi
\ifx \showLCCN     \undefined \def \showLCCN      #1{\unskip}     \fi
\ifx \shownote     \undefined \def \shownote      #1{#1}          \fi
\ifx \showarticletitle \undefined \def \showarticletitle #1{#1}   \fi
\ifx \showURL      \undefined \def \showURL       {\relax}        \fi
\providecommand\bibfield[2]{#2}
\providecommand\bibinfo[2]{#2}
\providecommand\natexlab[1]{#1}
\providecommand\showeprint[2][]{arXiv:#2}

\bibitem[Achiam et~al\mbox{.}(2023)]%
        {openai-gpt4}
\bibfield{author}{\bibinfo{person}{Josh Achiam}, \bibinfo{person}{Steven Adler}, \bibinfo{person}{Sandhini Agarwal}, \bibinfo{person}{Lama Ahmad}, \bibinfo{person}{Ilge Akkaya}, \bibinfo{person}{Florencia~Leoni Aleman}, \bibinfo{person}{Diogo Almeida}, \bibinfo{person}{Janko Altenschmidt}, \bibinfo{person}{Sam Altman}, \bibinfo{person}{Shyamal Anadkat}, {et~al\mbox{.}}} \bibinfo{year}{2023}\natexlab{}.
\newblock \showarticletitle{Gpt-4 technical report}.
\newblock \bibinfo{journal}{\emph{arXiv preprint arXiv:2303.08774}} (\bibinfo{year}{2023}).
\newblock


\bibitem[Ahmetovic et~al\mbox{.}(2020)]%
        {ahmetovic2020recog}
\bibfield{author}{\bibinfo{person}{Dragan Ahmetovic}, \bibinfo{person}{Daisuke Sato}, \bibinfo{person}{Uran Oh}, \bibinfo{person}{Tatsuya Ishihara}, \bibinfo{person}{Kris Kitani}, {and} \bibinfo{person}{Chieko Asakawa}.} \bibinfo{year}{2020}\natexlab{}.
\newblock \showarticletitle{ReCog: Supporting Blind People in Recognizing Personal Objects}. In \bibinfo{booktitle}{\emph{Proceedings of the 2020 CHI Conference on Human Factors in Computing Systems}} (Honolulu, HI, USA) \emph{(\bibinfo{series}{CHI '20})}. \bibinfo{publisher}{ACM}, \bibinfo{address}{New York, NY, USA}, \bibinfo{pages}{1–12}.
\newblock
\showISBNx{9781450367080}
\urldef\tempurl%
\url{https://doi.org/10.1145/3313831.3376143}
\showDOI{\tempurl}


\bibitem[{Aira Company}(2024)]%
        {aira}
\bibfield{author}{\bibinfo{person}{{Aira Company}}.} \bibinfo{year}{2024}\natexlab{}.
\newblock \bibinfo{title}{Aira: Visual assistance for blind and low-vision individuals}.
\newblock
\newblock
\urldef\tempurl%
\url{https://www.aira.io}
\showURL{%
\tempurl}
\newblock
\shownote{Retrieved July 5, 2024}.


\bibitem[Awad et~al\mbox{.}(2018)]%
        {milios2018intelligenteye}
\bibfield{author}{\bibinfo{person}{Milios Awad}, \bibinfo{person}{Jad~El Haddad}, \bibinfo{person}{Edgar Khneisser}, \bibinfo{person}{Tarek Mahmoud}, \bibinfo{person}{Elias Yaacoub}, {and} \bibinfo{person}{Mohammad Malli}.} \bibinfo{year}{2018}\natexlab{}.
\newblock \showarticletitle{Intelligent eye: A mobile application for assisting blind people}. In \bibinfo{booktitle}{\emph{Proceedings of the IEEE Middle East and North Africa Communications Conference (MENACOMM)}}. \bibinfo{pages}{1--6}.
\newblock
\urldef\tempurl%
\url{https://doi.org/10.1109/MENACOMM.2018.8371005}
\showDOI{\tempurl}


\bibitem[Aydemir et~al\mbox{.}(2013)]%
        {aydemir2013active}
\bibfield{author}{\bibinfo{person}{Alper Aydemir}, \bibinfo{person}{Andrzej Pronobis}, \bibinfo{person}{Moritz Göbelbecker}, {and} \bibinfo{person}{Patric Jensfelt}.} \bibinfo{year}{2013}\natexlab{}.
\newblock \showarticletitle{Active Visual Object Search in Unknown Environments Using Uncertain Semantics}.
\newblock \bibinfo{journal}{\emph{IEEE Transactions on Robotics}} \bibinfo{volume}{29}, \bibinfo{number}{4} (\bibinfo{year}{2013}), \bibinfo{pages}{986--1002}.
\newblock
\urldef\tempurl%
\url{https://doi.org/10.1109/TRO.2013.2256686}
\showDOI{\tempurl}


\bibitem[Bala et~al\mbox{.}(2023)]%
        {Bala2023}
\bibfield{author}{\bibinfo{person}{Myneni~Madhu Bala}, \bibinfo{person}{D.~N. Vasundhara}, \bibinfo{person}{Akkineni Haritha}, {and} \bibinfo{person}{CH. V. K. N. S.~N. Moorthy}.} \bibinfo{year}{2023}\natexlab{}.
\newblock \showarticletitle{Design, development and performance analysis of cognitive assisting aid with multi sensor fused navigation for visually impaired people}.
\newblock \bibinfo{journal}{\emph{Journal of Big Data}} \bibinfo{volume}{10}, \bibinfo{number}{1} (\bibinfo{year}{2023}), \bibinfo{pages}{21}.
\newblock
\urldef\tempurl%
\url{https://doi.org/10.1186/s40537-023-00689-5}
\showDOI{\tempurl}


\bibitem[Bennett et~al\mbox{.}(2018)]%
        {bennett2018independence}
\bibfield{author}{\bibinfo{person}{Cynthia~L. Bennett}, \bibinfo{person}{Erin Brady}, {and} \bibinfo{person}{Stacy~M. Branham}.} \bibinfo{year}{2018}\natexlab{}.
\newblock \showarticletitle{Interdependence as a Frame for Assistive Technology Research and Design}. In \bibinfo{booktitle}{\emph{Proceedings of the 20th ASSETS}} (Galway, Ireland) \emph{(\bibinfo{series}{ASSETS '18})}. \bibinfo{publisher}{Association for Computing Machinery}, \bibinfo{address}{New York, NY, USA}, \bibinfo{pages}{161–173}.
\newblock
\showISBNx{9781450356503}
\urldef\tempurl%
\url{https://doi.org/10.1145/3234695.3236348}
\showDOI{\tempurl}


\bibitem[Bigham et~al\mbox{.}(2010)]%
        {bigham2010locateit}
\bibfield{author}{\bibinfo{person}{Jeffrey~P. Bigham}, \bibinfo{person}{Chandrika Jayant}, \bibinfo{person}{Andrew Miller}, \bibinfo{person}{Brandyn White}, {and} \bibinfo{person}{Tom Yeh}.} \bibinfo{year}{2010}\natexlab{}.
\newblock \showarticletitle{VizWiz::LocateIt - enabling blind people to locate objects in their environment}. In \bibinfo{booktitle}{\emph{2010 IEEE Computer Society Conference on Computer Vision and Pattern Recognition - Workshops}}. \bibinfo{pages}{65--72}.
\newblock
\urldef\tempurl%
\url{https://doi.org/10.1109/CVPRW.2010.5543821}
\showDOI{\tempurl}


\bibitem[Boldu et~al\mbox{.}(2020)]%
        {boldu2020aisee}
\bibfield{author}{\bibinfo{person}{Roger Boldu}, \bibinfo{person}{Denys~J.C. Matthies}, \bibinfo{person}{Haimo Zhang}, {and} \bibinfo{person}{Suranga Nanayakkara}.} \bibinfo{year}{2020}\natexlab{}.
\newblock \showarticletitle{AiSee: An Assistive Wearable Device to Support Visually Impaired Grocery Shoppers}.
\newblock \bibinfo{journal}{\emph{Proceedings of the ACM on Interactive, Mobile, Wearable and Ubiquitous Technologies}} \bibinfo{volume}{4}, \bibinfo{number}{4}, Article \bibinfo{articleno}{119} (\bibinfo{date}{dec} \bibinfo{year}{2020}), \bibinfo{numpages}{25}~pages.
\newblock
\urldef\tempurl%
\url{https://doi.org/10.1145/3432196}
\showDOI{\tempurl}


\bibitem[Brady et~al\mbox{.}(2013)]%
        {brady2013visual}
\bibfield{author}{\bibinfo{person}{Erin Brady}, \bibinfo{person}{Meredith~Ringel Morris}, \bibinfo{person}{Yu Zhong}, \bibinfo{person}{Samuel White}, {and} \bibinfo{person}{Jeffrey~P Bigham}.} \bibinfo{year}{2013}\natexlab{}.
\newblock \showarticletitle{Visual challenges in the everyday lives of blind people}. In \bibinfo{booktitle}{\emph{Proceedings of the SIGCHI conference on human factors in computing systems}}. \bibinfo{pages}{2117--2126}.
\newblock


\bibitem[Cao et~al\mbox{.}(2024)]%
        {cao2024cognav}
\bibfield{author}{\bibinfo{person}{Yihan Cao}, \bibinfo{person}{Jiazhao Zhang}, \bibinfo{person}{Zhinan Yu}, \bibinfo{person}{Shuzhen Liu}, \bibinfo{person}{Zheng Qin}, \bibinfo{person}{Qin Zou}, \bibinfo{person}{Bo Du}, {and} \bibinfo{person}{Kai Xu}.} \bibinfo{year}{2024}\natexlab{}.
\newblock \showarticletitle{CogNav: Cognitive Process Modeling for Object Goal Navigation with LLMs}.
\newblock \bibinfo{journal}{\emph{arXiv preprint arXiv:2412.10439}} (\bibinfo{year}{2024}).
\newblock


\bibitem[Carrol(1995)]%
        {carrol1995scenario}
\bibfield{author}{\bibinfo{person}{JOHN~M Carrol}.} \bibinfo{year}{1995}\natexlab{}.
\newblock \showarticletitle{Scenario-based design: envisioning work and technology in system development}.
\newblock \bibinfo{journal}{\emph{NY John Wiley \& Sons, Inc}} (\bibinfo{year}{1995}).
\newblock


\bibitem[Chang et~al\mbox{.}(2024)]%
        {chang2024worldscribe}
\bibfield{author}{\bibinfo{person}{Ruei-Che Chang}, \bibinfo{person}{Yuxuan Liu}, {and} \bibinfo{person}{Anhong Guo}.} \bibinfo{year}{2024}\natexlab{}.
\newblock \showarticletitle{WorldScribe: Towards Context-Aware Live Visual Descriptions}. In \bibinfo{booktitle}{\emph{Proceedings of the 37th Annual ACM Symposium on User Interface Software and Technology}}. \bibinfo{pages}{1--18}.
\newblock


\bibitem[Chaplot et~al\mbox{.}(2020)]%
        {chaplot2020object}
\bibfield{author}{\bibinfo{person}{Devendra~Singh Chaplot}, \bibinfo{person}{Dhiraj~Prakashchand Gandhi}, \bibinfo{person}{Abhinav Gupta}, {and} \bibinfo{person}{Russ~R Salakhutdinov}.} \bibinfo{year}{2020}\natexlab{}.
\newblock \showarticletitle{Object goal navigation using goal-oriented semantic exploration}.
\newblock \bibinfo{journal}{\emph{Advances in Neural Information Processing Systems}}  \bibinfo{volume}{33} (\bibinfo{year}{2020}), \bibinfo{pages}{4247--4258}.
\newblock


\bibitem[Cheng et~al\mbox{.}(2024)]%
        {Cheng2024YOLOWorld}
\bibfield{author}{\bibinfo{person}{Tianheng Cheng}, \bibinfo{person}{Lin Song}, \bibinfo{person}{Yixiao Ge}, \bibinfo{person}{Wenyu Liu}, \bibinfo{person}{Xinggang Wang}, {and} \bibinfo{person}{Ying Shan}.} \bibinfo{year}{2024}\natexlab{}.
\newblock \showarticletitle{YOLO-World: Real-Time Open-Vocabulary Object Detection}. In \bibinfo{booktitle}{\emph{CVPR}}. IEEE, \bibinfo{pages}{16901--16911}.
\newblock


\bibitem[Cockburn(2000)]%
        {cockburn2000usecase}
\bibfield{author}{\bibinfo{person}{Alistair Cockburn}.} \bibinfo{year}{2000}\natexlab{}.
\newblock \bibinfo{booktitle}{\emph{Writing Effective Use Cases} (\bibinfo{edition}{1st} ed.)}.
\newblock \bibinfo{publisher}{Addison-Wesley Longman Publishing Co., Inc.}, \bibinfo{address}{USA}.
\newblock
\showISBNx{0201702258}


\bibitem[Constantinescu et~al\mbox{.}(2020a)]%
        {constantinescu2020blind}
\bibfield{author}{\bibinfo{person}{Angela Constantinescu}, \bibinfo{person}{Karin M\"{u}ller}, \bibinfo{person}{Monica Haurilet}, \bibinfo{person}{Vanessa Petrausch}, {and} \bibinfo{person}{Rainer Stiefelhagen}.} \bibinfo{year}{2020}\natexlab{a}.
\newblock \showarticletitle{Bring the Environment to Life: A Sonification Module for People with Visual Impairments to Improve Situation Awareness}. In \bibinfo{booktitle}{\emph{Proceedings of the International Conference on Multimodal Interaction}} (Virtual Event, Netherlands) \emph{(\bibinfo{series}{ICMI '20})}. \bibinfo{publisher}{ACM}, \bibinfo{address}{New York, NY, USA}, \bibinfo{pages}{50–59}.
\newblock
\showISBNx{9781450375818}
\urldef\tempurl%
\url{https://doi.org/10.1145/3382507.3418874}
\showDOI{\tempurl}


\bibitem[Constantinescu et~al\mbox{.}(2020b)]%
        {constantinescu2020bring}
\bibfield{author}{\bibinfo{person}{Angela Constantinescu}, \bibinfo{person}{Karin M\"{u}ller}, \bibinfo{person}{Monica Haurilet}, \bibinfo{person}{Vanessa Petrausch}, {and} \bibinfo{person}{Rainer Stiefelhagen}.} \bibinfo{year}{2020}\natexlab{b}.
\newblock \showarticletitle{Bring the Environment to Life: A Sonification Module for People with Visual Impairments to Improve Situation Awareness}. In \bibinfo{booktitle}{\emph{Proceedings of the 2020 International Conference on Multimodal Interaction}} (Virtual Event, Netherlands) \emph{(\bibinfo{series}{ICMI '20})}. \bibinfo{publisher}{Association for Computing Machinery}, \bibinfo{address}{New York, NY, USA}, \bibinfo{pages}{50–59}.
\newblock
\showISBNx{9781450375818}
\urldef\tempurl%
\url{https://doi.org/10.1145/3382507.3418874}
\showDOI{\tempurl}


\bibitem[Constantinescu et~al\mbox{.}(2022)]%
        {Constantinescu2022listening}
\bibfield{author}{\bibinfo{person}{Angela Constantinescu}, \bibinfo{person}{Eva-Maria Neumann}, \bibinfo{person}{Karin M\"{u}ller}, \bibinfo{person}{Gerhard Jaworek}, {and} \bibinfo{person}{Rainer Stiefelhagen}.} \bibinfo{year}{2022}\natexlab{}.
\newblock \showarticletitle{Listening First: Egocentric Textual Descriptions of Indoor Spaces for People with Blindness}. In \bibinfo{booktitle}{\emph{Proceedings of the International Conference on Computers Helping People with Special Needs}} (Milan, Italy). \bibinfo{publisher}{Springer-Verlag}, \bibinfo{address}{Berlin, Heidelberg}, \bibinfo{pages}{241–249}.
\newblock
\showISBNx{978-3-031-08647-2}
\urldef\tempurl%
\url{https://doi.org/10.1007/978-3-031-08648-9_28}
\showDOI{\tempurl}


\bibitem[Corporation(2024)]%
        {seeingAI}
\bibfield{author}{\bibinfo{person}{Microsoft Corporation}.} \bibinfo{year}{2024}\natexlab{}.
\newblock \bibinfo{title}{Seeing AI}.
\newblock
\newblock
\urldef\tempurl%
\url{https://www.microsoft.com/en-us/ai/seeing-ai}
\showURL{%
\tempurl}
\newblock
\shownote{Accessed: 2024-09-01}.


\bibitem[Cummings and Bauchwitz(2024)]%
        {cummings2024unreliable}
\bibfield{author}{\bibinfo{person}{Mary~L Cummings} {and} \bibinfo{person}{Ben Bauchwitz}.} \bibinfo{year}{2024}\natexlab{}.
\newblock \showarticletitle{Unreliable Pedestrian Detection and Driver Alerting in Intelligent Vehicles}.
\newblock \bibinfo{journal}{\emph{IEEE Transactions on Intelligent Vehicles}} (\bibinfo{year}{2024}).
\newblock


\bibitem[Duh et~al\mbox{.}(2020)]%
        {duh2020veye}
\bibfield{author}{\bibinfo{person}{Ping-Jung Duh}, \bibinfo{person}{Yu-Cheng Sung}, \bibinfo{person}{Liang-Yu~Fan Chiang}, \bibinfo{person}{Yung-Ju Chang}, {and} \bibinfo{person}{Kuan-Wen Chen}.} \bibinfo{year}{2020}\natexlab{}.
\newblock \showarticletitle{V-eye: A vision-based navigation system for the visually impaired}.
\newblock \bibinfo{journal}{\emph{IEEE Transactions on Multimedia}}  \bibinfo{volume}{23} (\bibinfo{year}{2020}), \bibinfo{pages}{1567--1580}.
\newblock


\bibitem[Eyes(2023)]%
        {BeMyAI2023}
\bibfield{author}{\bibinfo{person}{Be~My Eyes}.} \bibinfo{year}{2023}\natexlab{}.
\newblock \bibinfo{title}{{Introducing Be My AI}}.
\newblock \bibinfo{howpublished}{\url{https://www.bemyeyes.com/blog/introducing-be-my-ai/}}.
\newblock
\newblock
\shownote{[Online; accessed 3-Apr-2025]}.


\bibitem[Eyes(2024)]%
        {bemyeyes}
\bibfield{author}{\bibinfo{person}{Be~My Eyes}.} \bibinfo{year}{2024}\natexlab{}.
\newblock \bibinfo{title}{{Be My Eyes App}}.
\newblock \bibinfo{howpublished}{\url{https://www.bemyeyes.com/}}.
\newblock
\newblock
\shownote{Accessed: 2024-06-14}.


\bibitem[Fereday and Muir-Cochrane(2006)]%
        {FeredayMuirCochrane2006}
\bibfield{author}{\bibinfo{person}{Jennifer Fereday} {and} \bibinfo{person}{Eimear Muir-Cochrane}.} \bibinfo{year}{2006}\natexlab{}.
\newblock \showarticletitle{Demonstrating Rigor Using Thematic Analysis: A Hybrid Approach of Inductive and Deductive Coding and Theme Development}.
\newblock \bibinfo{journal}{\emph{International Journal of Qualitative Methods}} \bibinfo{volume}{5}, \bibinfo{number}{1} (\bibinfo{year}{2006}), \bibinfo{pages}{80--92}.
\newblock
\urldef\tempurl%
\url{https://doi.org/10.1177/160940690600500107}
\showDOI{\tempurl}


\bibitem[for~the Blind(2024)]%
        {perkinsAltText}
\bibfield{author}{\bibinfo{person}{Perkins~School for~the Blind}.} \bibinfo{year}{2024}\natexlab{}.
\newblock \bibinfo{title}{How to Write Alt Text and Image Descriptions for the Visually Impaired}.
\newblock
\newblock
\urldef\tempurl%
\url{https://www.perkins.org/resource/how-write-alt-text-and-image-descriptions-visually-impaired/}
\showURL{%
\tempurl}
\newblock
\shownote{Accessed: 2024-06-17}.


\bibitem[Gamage et~al\mbox{.}(2023)]%
        {bhanuka2023want}
\bibfield{author}{\bibinfo{person}{Bhanuka Gamage}, \bibinfo{person}{Thanh-Toan Do}, \bibinfo{person}{Nicholas Seow~Chiang Price}, \bibinfo{person}{Arthur Lowery}, {and} \bibinfo{person}{Kim Marriott}.} \bibinfo{year}{2023}\natexlab{}.
\newblock \showarticletitle{What do Blind and Low-Vision People Really Want from Assistive Smart Devices? Comparison of the Literature with a Focus Study}. In \bibinfo{booktitle}{\emph{Proceedings of the 25th ASSETS}} (New York, NY, USA) \emph{(\bibinfo{series}{ASSETS '23})}. \bibinfo{publisher}{Association for Computing Machinery}, \bibinfo{address}{New York, NY, USA}, Article \bibinfo{articleno}{30}, \bibinfo{numpages}{21}~pages.
\newblock
\showISBNx{9798400702204}
\urldef\tempurl%
\url{https://doi.org/10.1145/3597638.3608955}
\showDOI{\tempurl}


\bibitem[Golledge(1999)]%
        {golledge1999wayfinding}
\bibfield{editor}{\bibinfo{person}{Reginald~G. Golledge}} (Ed.). \bibinfo{year}{1999}\natexlab{}.
\newblock \bibinfo{booktitle}{\emph{Wayfinding Behavior: Cognitive Mapping and Other Spatial Processes}}.
\newblock \bibinfo{publisher}{Johns Hopkins University Press}, \bibinfo{address}{Baltimore, MD, USA}.
\newblock
\showISBNx{978-0801859933}


\bibitem[Gonzalez~Penuela et~al\mbox{.}(2024)]%
        {Penuela2024usecase}
\bibfield{author}{\bibinfo{person}{Ricardo~E Gonzalez~Penuela}, \bibinfo{person}{Jazmin Collins}, \bibinfo{person}{Cynthia Bennett}, {and} \bibinfo{person}{Shiri Azenkot}.} \bibinfo{year}{2024}\natexlab{}.
\newblock \showarticletitle{Investigating Use Cases of AI-Powered Scene Description Applications for Blind and Low Vision People}. In \bibinfo{booktitle}{\emph{Proceedings of the CHI Conference on Human Factors in Computing Systems}} (Honolulu, HI, USA) \emph{(\bibinfo{series}{CHI '24})}. \bibinfo{publisher}{ACM}, \bibinfo{address}{New York, NY, USA}, Article \bibinfo{articleno}{901}, \bibinfo{numpages}{21}~pages.
\newblock
\showISBNx{9798400703300}
\urldef\tempurl%
\url{https://doi.org/10.1145/3613904.3642211}
\showDOI{\tempurl}


\bibitem[Google(2024)]%
        {lookout}
\bibfield{author}{\bibinfo{person}{Google}.} \bibinfo{year}{2024}\natexlab{}.
\newblock \bibinfo{title}{{Lookout}}.
\newblock \bibinfo{howpublished}{\url{https://play.google.com/store/apps/details?id=com.google.android.apps.accessibility.reveal}}.
\newblock
\newblock
\shownote{Accessed: 2024-06-14}.


\bibitem[Gurari et~al\mbox{.}(2018)]%
        {gurari2018vizwiz}
\bibfield{author}{\bibinfo{person}{Danna Gurari}, \bibinfo{person}{Qing Li}, \bibinfo{person}{Abigale~J Stangl}, \bibinfo{person}{Anhong Guo}, \bibinfo{person}{Chi Lin}, \bibinfo{person}{Kristen Grauman}, \bibinfo{person}{Jiebo Luo}, {and} \bibinfo{person}{Jeffrey~P Bigham}.} \bibinfo{year}{2018}\natexlab{}.
\newblock \showarticletitle{VizWiz Grand Challenge: Answering Visual Questions From Blind People}. In \bibinfo{booktitle}{\emph{CVPR}}. IEEE, \bibinfo{pages}{3608--3617}.
\newblock


\bibitem[Hart(2006)]%
        {hart2006nasa}
\bibfield{author}{\bibinfo{person}{Sandra~G Hart}.} \bibinfo{year}{2006}\natexlab{}.
\newblock \showarticletitle{NASA-task load index (NASA-TLX); 20 years later}. In \bibinfo{booktitle}{\emph{Proceedings of the human factors and ergonomics society annual meeting}}, Vol.~\bibinfo{volume}{50}. Sage publications Sage CA: Los Angeles, CA, \bibinfo{pages}{904--908}.
\newblock


\bibitem[Hersh(2022)]%
        {marion2022wearable}
\bibfield{author}{\bibinfo{person}{Marion Hersh}.} \bibinfo{year}{2022}\natexlab{}.
\newblock \showarticletitle{Wearable Travel Aids for Blind and Partially Sighted People: A Review with a Focus on Design Issues}.
\newblock \bibinfo{journal}{\emph{Sensors}} \bibinfo{volume}{22}, \bibinfo{number}{14} (\bibinfo{year}{2022}).
\newblock
\showISSN{1424-8220}
\urldef\tempurl%
\url{https://doi.org/10.3390/s22145454}
\showDOI{\tempurl}


\bibitem[Herskovitz et~al\mbox{.}(2023)]%
        {herskovitz2023diy}
\bibfield{author}{\bibinfo{person}{Jaylin Herskovitz}, \bibinfo{person}{Andi Xu}, \bibinfo{person}{Rahaf Alharbi}, {and} \bibinfo{person}{Anhong Guo}.} \bibinfo{year}{2023}\natexlab{}.
\newblock \showarticletitle{Hacking, Switching, Combining: Understanding and Supporting DIY Assistive Technology Design by Blind People}. In \bibinfo{booktitle}{\emph{Proceedings of the 2023 CHI Conference on Human Factors in Computing Systems}} (Hamburg, Germany) \emph{(\bibinfo{series}{CHI '23})}. \bibinfo{publisher}{Association for Computing Machinery}, \bibinfo{address}{New York, NY, USA}, Article \bibinfo{articleno}{57}, \bibinfo{numpages}{17}~pages.
\newblock
\showISBNx{9781450394215}
\urldef\tempurl%
\url{https://doi.org/10.1145/3544548.3581249}
\showDOI{\tempurl}


\bibitem[Herskovitz et~al\mbox{.}(2024)]%
        {herskovitz2024programally}
\bibfield{author}{\bibinfo{person}{Jaylin Herskovitz}, \bibinfo{person}{Andi Xu}, \bibinfo{person}{Rahaf Alharbi}, {and} \bibinfo{person}{Anhong Guo}.} \bibinfo{year}{2024}\natexlab{}.
\newblock \showarticletitle{ProgramAlly: Creating Custom Visual Access Programs via Multi-Modal End-User Programming}. In \bibinfo{booktitle}{\emph{Proceedings of the 37th Annual ACM Symposium on User Interface Software and Technology}}. \bibinfo{pages}{1--15}.
\newblock


\bibitem[Hong and Kacorri(2024)]%
        {hong2024understand}
\bibfield{author}{\bibinfo{person}{Jonggi Hong} {and} \bibinfo{person}{Hernisa Kacorri}.} \bibinfo{year}{2024}\natexlab{}.
\newblock \showarticletitle{Understanding How Blind Users Handle Object Recognition Errors: Strategies and Challenges}. In \bibinfo{booktitle}{\emph{Proceedings of the 26th ASSETS}} (St. John's, NL, Canada) \emph{(\bibinfo{series}{ASSETS '24})}. \bibinfo{publisher}{Association for Computing Machinery}, \bibinfo{address}{New York, NY, USA}, Article \bibinfo{articleno}{63}, \bibinfo{numpages}{15}~pages.
\newblock
\showISBNx{9798400706776}
\urldef\tempurl%
\url{https://doi.org/10.1145/3663548.3675635}
\showDOI{\tempurl}


\bibitem[Hong et~al\mbox{.}(2024)]%
        {hong2024spvinet}
\bibfield{author}{\bibinfo{person}{Kaipeng Hong}, \bibinfo{person}{Weiqin He}, \bibinfo{person}{Hui Tang}, \bibinfo{person}{Xing Zhang}, \bibinfo{person}{Qingquan Li}, {and} \bibinfo{person}{Baoding Zhou}.} \bibinfo{year}{2024}\natexlab{}.
\newblock \showarticletitle{SPVINet: A Lightweight Multitask Learning Network for Assisting Visually Impaired People in Multiscene Perception}.
\newblock \bibinfo{journal}{\emph{IEEE Internet of Things Journal}} (\bibinfo{year}{2024}).
\newblock


\bibitem[Hu et~al\mbox{.}(2022)]%
        {hu2022stereopilot}
\bibfield{author}{\bibinfo{person}{Xuhui Hu}, \bibinfo{person}{Aiguo Song}, \bibinfo{person}{Zhikai Wei}, {and} \bibinfo{person}{Hong Zeng}.} \bibinfo{year}{2022}\natexlab{}.
\newblock \showarticletitle{Stereopilot: A wearable target location system for blind and visually impaired using spatial audio rendering}.
\newblock \bibinfo{journal}{\emph{IEEE Transactions on Neural Systems and Rehabilitation Engineering}}  \bibinfo{volume}{30} (\bibinfo{year}{2022}), \bibinfo{pages}{1621--1630}.
\newblock


\bibitem[Islam et~al\mbox{.}(2023)]%
        {ISLAM2023e16924}
\bibfield{author}{\bibinfo{person}{Raihan~Bin Islam}, \bibinfo{person}{Samiha Akhter}, \bibinfo{person}{Faria Iqbal}, \bibinfo{person}{Md. {Saif Ur Rahman}}, {and} \bibinfo{person}{Riasat Khan}.} \bibinfo{year}{2023}\natexlab{}.
\newblock \showarticletitle{Deep learning based object detection and surrounding environment description for visually impaired people}.
\newblock \bibinfo{journal}{\emph{Heliyon}} \bibinfo{volume}{9}, \bibinfo{number}{6} (\bibinfo{year}{2023}), \bibinfo{pages}{e16924}.
\newblock
\showISSN{2405-8440}
\urldef\tempurl%
\url{https://doi.org/10.1016/j.heliyon.2023.e16924}
\showDOI{\tempurl}


\bibitem[Jain et~al\mbox{.}(2023a)]%
        {jain2023towards}
\bibfield{author}{\bibinfo{person}{Gaurav Jain}, \bibinfo{person}{Basel Hindi}, \bibinfo{person}{Mingyu Xie}, \bibinfo{person}{Zihao Zhang}, \bibinfo{person}{Koushik Srinivasula}, \bibinfo{person}{Mahshid Ghasemi}, \bibinfo{person}{Daniel Weiner}, \bibinfo{person}{Xin Yi~Therese Xu}, \bibinfo{person}{Sophie~Ana Paris}, \bibinfo{person}{Chloe Tedjo}, \bibinfo{person}{Josh Bassin}, \bibinfo{person}{Michael Malcolm}, \bibinfo{person}{Mehmet Turkcan}, \bibinfo{person}{Javad Ghaderi}, \bibinfo{person}{Zoran Kostic}, \bibinfo{person}{Gil Zussman}, {and} \bibinfo{person}{Brian~A. Smith}.} \bibinfo{year}{2023}\natexlab{a}.
\newblock \showarticletitle{Towards Street Camera-based Outdoor Navigation for Blind Pedestrians}. In \bibinfo{booktitle}{\emph{Proceedings of the 25th ASSETS}} (New York, NY, USA) \emph{(\bibinfo{series}{ASSETS '23})}. \bibinfo{publisher}{ACM}, \bibinfo{address}{New York, NY, USA}, Article \bibinfo{articleno}{77}, \bibinfo{numpages}{6}~pages.
\newblock
\showISBNx{9798400702204}
\urldef\tempurl%
\url{https://doi.org/10.1145/3597638.3614498}
\showDOI{\tempurl}


\bibitem[Jain et~al\mbox{.}(2023b)]%
        {jain2023want}
\bibfield{author}{\bibinfo{person}{Gaurav Jain}, \bibinfo{person}{Yuanyang Teng}, \bibinfo{person}{Dong~Heon Cho}, \bibinfo{person}{Yunhao Xing}, \bibinfo{person}{Maryam Aziz}, {and} \bibinfo{person}{Brian~A Smith}.} \bibinfo{year}{2023}\natexlab{b}.
\newblock \showarticletitle{" I Want to Figure Things Out": Supporting Exploration in Navigation for People with Visual Impairments}.
\newblock \bibinfo{journal}{\emph{Proceedings of the ACM on Human-Computer Interaction}} \bibinfo{volume}{7}, \bibinfo{number}{CSCW1} (\bibinfo{year}{2023}), \bibinfo{pages}{1--28}.
\newblock


\bibitem[Jones and Kenward(2014)]%
        {jones2014crossover}
\bibfield{author}{\bibinfo{person}{Byron Jones} {and} \bibinfo{person}{Michael~G. Kenward}.} \bibinfo{year}{2014}\natexlab{}.
\newblock \bibinfo{booktitle}{\emph{Design and Analysis of Cross-Over Trials} (\bibinfo{edition}{3} ed.)}.
\newblock \bibinfo{publisher}{CRC Press}.
\newblock


\bibitem[Kacorri et~al\mbox{.}(2017)]%
        {kacorri2017classifier}
\bibfield{author}{\bibinfo{person}{Hernisa Kacorri}, \bibinfo{person}{Kris~M. Kitani}, \bibinfo{person}{Jeffrey~P. Bigham}, {and} \bibinfo{person}{Chieko Asakawa}.} \bibinfo{year}{2017}\natexlab{}.
\newblock \showarticletitle{People with Visual Impairment Training Personal Object Recognizers: Feasibility and Challenges}. In \bibinfo{booktitle}{\emph{Proceedings of the 2017 CHI Conference on Human Factors in Computing Systems}} (Denver, Colorado, USA) \emph{(\bibinfo{series}{CHI '17})}. \bibinfo{publisher}{ACM}, \bibinfo{address}{New York, NY, USA}, \bibinfo{pages}{5839–5849}.
\newblock
\showISBNx{9781450346559}
\urldef\tempurl%
\url{https://doi.org/10.1145/3025453.3025899}
\showDOI{\tempurl}


\bibitem[Kamikubo et~al\mbox{.}(2020)]%
        {kamikubo2020support}
\bibfield{author}{\bibinfo{person}{Rie Kamikubo}, \bibinfo{person}{Naoya Kato}, \bibinfo{person}{Keita Higuchi}, \bibinfo{person}{Ryo Yonetani}, {and} \bibinfo{person}{Yoichi Sato}.} \bibinfo{year}{2020}\natexlab{}.
\newblock \showarticletitle{Support Strategies for Remote Guides in Assisting People with Visual Impairments for Effective Indoor Navigation}. In \bibinfo{booktitle}{\emph{Proceedings of the 2020 CHI Conference on Human Factors in Computing Systems}} (Honolulu, HI, USA) \emph{(\bibinfo{series}{CHI '20})}. \bibinfo{publisher}{Association for Computing Machinery}, \bibinfo{address}{New York, NY, USA}, \bibinfo{pages}{1–12}.
\newblock
\showISBNx{9781450367080}
\urldef\tempurl%
\url{https://doi.org/10.1145/3313831.3376823}
\showDOI{\tempurl}


\bibitem[Kane et~al\mbox{.}(2020)]%
        {kane2020sense}
\bibfield{author}{\bibinfo{person}{Shaun~K. Kane}, \bibinfo{person}{Anhong Guo}, {and} \bibinfo{person}{Meredith~Ringel Morris}.} \bibinfo{year}{2020}\natexlab{}.
\newblock \showarticletitle{Sense and Accessibility: Understanding People with Physical Disabilities’ Experiences with Sensing Systems}. In \bibinfo{booktitle}{\emph{Proceedings of the 22nd ASSETS}} (Virtual Event, Greece) \emph{(\bibinfo{series}{ASSETS '20})}. \bibinfo{publisher}{Association for Computing Machinery}, \bibinfo{address}{New York, NY, USA}, Article \bibinfo{articleno}{42}, \bibinfo{numpages}{14}~pages.
\newblock
\showISBNx{9781450371032}
\urldef\tempurl%
\url{https://doi.org/10.1145/3373625.3416990}
\showDOI{\tempurl}


\bibitem[Kawaharazuka et~al\mbox{.}(2024)]%
        {kawaharazuka2024reflex}
\bibfield{author}{\bibinfo{person}{Kento Kawaharazuka}, \bibinfo{person}{Yoshiki Obinata}, \bibinfo{person}{Naoaki Kanazawa}, \bibinfo{person}{Naoto Tsukamoto}, \bibinfo{person}{Kei Okada}, {and} \bibinfo{person}{Masayuki Inaba}.} \bibinfo{year}{2024}\natexlab{}.
\newblock \showarticletitle{Reflex-based open-vocabulary navigation without prior knowledge using omnidirectional camera and multiple vision-language models}.
\newblock \bibinfo{journal}{\emph{Advanced Robotics}} \bibinfo{volume}{0}, \bibinfo{number}{0} (\bibinfo{year}{2024}), \bibinfo{pages}{1--11}.
\newblock
\urldef\tempurl%
\url{https://doi.org/10.1080/01691864.2024.2393409}
\showDOI{\tempurl}


\bibitem[Kirillov et~al\mbox{.}(2023)]%
        {kirillov2023segment}
\bibfield{author}{\bibinfo{person}{Alexander Kirillov}, \bibinfo{person}{Eric Mintun}, \bibinfo{person}{Nikhila Ravi}, \bibinfo{person}{Hanzi Mao}, \bibinfo{person}{Chloe Rolland}, \bibinfo{person}{Laura Gustafson}, \bibinfo{person}{Tete Xiao}, \bibinfo{person}{Spencer Whitehead}, \bibinfo{person}{Alexander~C Berg}, \bibinfo{person}{Wan-Yen Lo}, {et~al\mbox{.}}} \bibinfo{year}{2023}\natexlab{}.
\newblock \showarticletitle{Segment anything}. In \bibinfo{booktitle}{\emph{ICCV}}. IEEE, \bibinfo{pages}{3992--4003}.
\newblock


\bibitem[{KR Vision}({[n.\,d.]})]%
        {krvision}
\bibfield{author}{\bibinfo{person}{{KR Vision}}.} \bibinfo{year}{[n.\,d.]}\natexlab{}.
\newblock \bibinfo{title}{{KR Vision Website}}.
\newblock \bibinfo{howpublished}{\url{http://krvision.com.cn/home}}.
\newblock
\newblock
\shownote{Accessed: 2024-08-31}.


\bibitem[Kuribayashi et~al\mbox{.}(2025)]%
        {kuribayashi2025wanderguide}
\bibfield{author}{\bibinfo{person}{Masaki Kuribayashi}, \bibinfo{person}{Kohei Uehara}, \bibinfo{person}{Allan Wang}, \bibinfo{person}{Shigeo Morishima}, {and} \bibinfo{person}{Chieko Asakawa}.} \bibinfo{year}{2025}\natexlab{}.
\newblock \showarticletitle{WanderGuide: Indoor Map-less Robotic Guide for Exploration by Blind People}. In \bibinfo{booktitle}{\emph{Proceedings of the 2025 CHI Conference on Human Factors in Computing Systems (CHI '25)}}. \bibinfo{publisher}{ACM}, \bibinfo{address}{New York, NY, USA}.
\newblock
\urldef\tempurl%
\url{https://doi.org/10.48550/arXiv.2502.08906}
\showDOI{\tempurl}


\bibitem[Lee et~al\mbox{.}(2022a)]%
        {lee2022imageexplorer}
\bibfield{author}{\bibinfo{person}{Jaewook Lee}, \bibinfo{person}{Jaylin Herskovitz}, \bibinfo{person}{Yi-Hao Peng}, {and} \bibinfo{person}{Anhong Guo}.} \bibinfo{year}{2022}\natexlab{a}.
\newblock \showarticletitle{ImageExplorer: Multi-Layered Touch Exploration to Encourage Skepticism Towards Imperfect AI-Generated Image Captions}. In \bibinfo{booktitle}{\emph{Proceedings of the 2022 CHI Conference on Human Factors in Computing Systems}} (New Orleans, LA, USA) \emph{(\bibinfo{series}{CHI '22})}. \bibinfo{publisher}{ACM}, \bibinfo{address}{New York, NY, USA}, Article \bibinfo{articleno}{462}, \bibinfo{numpages}{15}~pages.
\newblock
\showISBNx{9781450391573}
\urldef\tempurl%
\url{https://doi.org/10.1145/3491102.3501966}
\showDOI{\tempurl}


\bibitem[Lee et~al\mbox{.}(2020)]%
        {lee2020emerging}
\bibfield{author}{\bibinfo{person}{Sooyeon Lee}, \bibinfo{person}{Madison Reddie}, \bibinfo{person}{Chun-Hua Tsai}, \bibinfo{person}{Jordan Beck}, \bibinfo{person}{Mary~Beth Rosson}, {and} \bibinfo{person}{John~M. Carroll}.} \bibinfo{year}{2020}\natexlab{}.
\newblock \showarticletitle{The Emerging Professional Practice of Remote Sighted Assistance for People with Visual Impairments}. In \bibinfo{booktitle}{\emph{Proceedings of the 2020 CHI Conference on Human Factors in Computing Systems}} (Honolulu, HI, USA) \emph{(\bibinfo{series}{CHI '20})}. \bibinfo{publisher}{Association for Computing Machinery}, \bibinfo{address}{New York, NY, USA}, \bibinfo{pages}{1–12}.
\newblock
\showISBNx{9781450367080}
\urldef\tempurl%
\url{https://doi.org/10.1145/3313831.3376591}
\showDOI{\tempurl}


\bibitem[Lee et~al\mbox{.}(2022b)]%
        {lee2022opportunity}
\bibfield{author}{\bibinfo{person}{Sooyeon Lee}, \bibinfo{person}{Rui Yu}, \bibinfo{person}{Jingyi Xie}, \bibinfo{person}{Syed~Masum Billah}, {and} \bibinfo{person}{John~M. Carroll}.} \bibinfo{year}{2022}\natexlab{b}.
\newblock \showarticletitle{Opportunities for Human-AI Collaboration in Remote Sighted Assistance}. In \bibinfo{booktitle}{\emph{Proceedings of the 27th International Conference on Intelligent User Interfaces}} (Helsinki, Finland) \emph{(\bibinfo{series}{IUI '22})}. \bibinfo{publisher}{Association for Computing Machinery}, \bibinfo{address}{New York, NY, USA}, \bibinfo{pages}{63–78}.
\newblock
\showISBNx{9781450391443}
\urldef\tempurl%
\url{https://doi.org/10.1145/3490099.3511113}
\showDOI{\tempurl}


\bibitem[Li et~al\mbox{.}(2023)]%
        {li2023multi}
\bibfield{author}{\bibinfo{person}{Guoxin Li}, \bibinfo{person}{Zhijun Li}, \bibinfo{person}{Haisheng Xia}, {and} \bibinfo{person}{Ying Feng}.} \bibinfo{year}{2023}\natexlab{}.
\newblock \showarticletitle{Multi-Sensory Visual-Auditory Fusion of Wearable Navigation Assistance for People with Impaired Vision}. In \bibinfo{booktitle}{\emph{Proceedings of 2023 IEEE International Conference on Systems, Man, and Cybernetics (SMC)}}. IEEE, \bibinfo{pages}{955--960}.
\newblock


\bibitem[Lin et~al\mbox{.}(2014)]%
        {lin2014microsoft}
\bibfield{author}{\bibinfo{person}{Tsung-Yi Lin}, \bibinfo{person}{Michael Maire}, \bibinfo{person}{Serge Belongie}, \bibinfo{person}{James Hays}, \bibinfo{person}{Pietro Perona}, \bibinfo{person}{Deva Ramanan}, \bibinfo{person}{Piotr Doll{\'a}r}, {and} \bibinfo{person}{C.~Lawrence Zitnick}.} \bibinfo{year}{2014}\natexlab{}.
\newblock \showarticletitle{Microsoft COCO: Common Objects in Context}. In \bibinfo{booktitle}{\emph{Computer Vision--ECCV 2014: 13th European Conference, Zurich, Switzerland, September 6-12, 2014, Proceedings, Part V 13}}, \bibfield{editor}{\bibinfo{person}{David Fleet}, \bibinfo{person}{Tomas Pajdla}, \bibinfo{person}{Bernt Schiele}, {and} \bibinfo{person}{Tinne Tuytelaars}} (Eds.). \bibinfo{publisher}{Springer International Publishing}, \bibinfo{address}{Cham}, \bibinfo{pages}{740--755}.
\newblock
\showISBNx{978-3-319-10602-1}


\bibitem[Liu et~al\mbox{.}(2021)]%
        {liu2021hida}
\bibfield{author}{\bibinfo{person}{Huayao Liu}, \bibinfo{person}{Ruiping Liu}, \bibinfo{person}{Kailun Yang}, \bibinfo{person}{Jiaming Zhang}, \bibinfo{person}{Kunyu Peng}, {and} \bibinfo{person}{Rainer Stiefelhagen}.} \bibinfo{year}{2021}\natexlab{}.
\newblock \showarticletitle{HIDA: Towards Holistic Indoor Understanding for the Visually Impaired via Semantic Instance Segmentation with a Wearable Solid-State LiDAR Sensor}. In \bibinfo{booktitle}{\emph{Proceedings of the IEEE/CVF International Conference on Computer Vision Workshops (ICCVW)}}. IEEE, \bibinfo{pages}{1780--1790}.
\newblock


\bibitem[Liu(2023)]%
        {liu2023realtime}
\bibfield{author}{\bibinfo{person}{Jenny Liu}.} \bibinfo{year}{2023}\natexlab{}.
\newblock \showarticletitle{Real-Time Machine Learning Based Object Detection and Recognition System for the Visually Impaired}. In \bibinfo{booktitle}{\emph{Proceedings of the 2023 Workshop on Advanced Multimedia Computing for Smart Manufacturing and Engineering}} (Ottawa ON, Canada) \emph{(\bibinfo{series}{AMC-SME '23})}. \bibinfo{publisher}{ACM}, \bibinfo{address}{New York, NY, USA}, \bibinfo{pages}{31–35}.
\newblock
\showISBNx{9798400702730}
\urldef\tempurl%
\url{https://doi.org/10.1145/3606042.3616454}
\showDOI{\tempurl}


\bibitem[Liu et~al\mbox{.}(2023)]%
        {liu2023opensu}
\bibfield{author}{\bibinfo{person}{Ruiping Liu}, \bibinfo{person}{Jiaming Zhang}, \bibinfo{person}{Kunyu Peng}, \bibinfo{person}{Junwei Zheng}, \bibinfo{person}{Ke Cao}, \bibinfo{person}{Yufan Chen}, \bibinfo{person}{Kailun Yang}, {and} \bibinfo{person}{Rainer Stiefelhagen}.} \bibinfo{year}{2023}\natexlab{}.
\newblock \showarticletitle{Open Scene Understanding: Grounded Situation Recognition Meets Segment Anything for Helping People with Visual Impairments}. In \bibinfo{booktitle}{\emph{Proceedings of the IEEE/CVF International Conference on Computer Vision Workshops (ICCVW)}}. IEEE, \bibinfo{pages}{1849--1859}.
\newblock


\bibitem[Martolini et~al\mbox{.}(2021)]%
        {martolini2021allocentric}
\bibfield{author}{\bibinfo{person}{Chiara Martolini}, \bibinfo{person}{Giulia Cappagli}, \bibinfo{person}{Elena Saligari}, \bibinfo{person}{Monica Gori}, {and} \bibinfo{person}{Sabrina Signorini}.} \bibinfo{year}{2021}\natexlab{}.
\newblock \showarticletitle{Allocentric spatial perception through vision and touch in sighted and blind children}.
\newblock \bibinfo{journal}{\emph{Journal of Experimental Child Psychology}}  \bibinfo{volume}{210} (\bibinfo{year}{2021}), \bibinfo{pages}{105195}.
\newblock
\showISSN{0022-0965}
\urldef\tempurl%
\url{https://doi.org/10.1016/j.jecp.2021.105195}
\showDOI{\tempurl}


\bibitem[Mathis and Sch{\"o}ning(2025)]%
        {mathis2025lifeinsight}
\bibfield{author}{\bibinfo{person}{Florian Mathis} {and} \bibinfo{person}{Johannes Sch{\"o}ning}.} \bibinfo{year}{2025}\natexlab{}.
\newblock \showarticletitle{LifeInsight: Design and Evaluation of an AI-Powered Assistive Wearable for Blind and Low Vision People Across Multiple Everyday Life Scenarios}. In \bibinfo{booktitle}{\emph{Proceedings of the 2025 CHI Conference on Human Factors in Computing Systems}}.
\newblock


\bibitem[M\"{u}ller et~al\mbox{.}(2022)]%
        {muller2022traveling}
\bibfield{author}{\bibinfo{person}{Karin M\"{u}ller}, \bibinfo{person}{Christin Engel}, \bibinfo{person}{Claudia Loitsch}, \bibinfo{person}{Rainer Stiefelhagen}, {and} \bibinfo{person}{Gerhard Weber}.} \bibinfo{year}{2022}\natexlab{}.
\newblock \showarticletitle{Traveling More Independently: A Study on the Diverse Needs and Challenges of People with Visual or Mobility Impairments in Unfamiliar Indoor Environments}.
\newblock \bibinfo{journal}{\emph{ACM Transactions on Accessible Computing}} \bibinfo{volume}{15}, \bibinfo{number}{2}, Article \bibinfo{articleno}{13} (\bibinfo{date}{may} \bibinfo{year}{2022}), \bibinfo{numpages}{44}~pages.
\newblock
\showISSN{1936-7228}
\urldef\tempurl%
\url{https://doi.org/10.1145/3514255}
\showDOI{\tempurl}


\bibitem[{netz-barrierefrei.de}(nd)]%
        {netz2025what}
\bibfield{author}{\bibinfo{person}{{netz-barrierefrei.de}}.} \bibinfo{year}{n.d.}\natexlab{}.
\newblock \bibinfo{title}{What is Blind Life?}
\newblock \bibinfo{howpublished}{\url{https://netz-barrierefrei.de/en/what-is-blind-life.html}}.
\newblock
\newblock
\shownote{[Online; accessed 05-April-2025]}.


\bibitem[{OpenAI}(2023)]%
        {OpenAIPromptEngineering}
\bibfield{author}{\bibinfo{person}{{OpenAI}}.} \bibinfo{year}{2023}\natexlab{}.
\newblock \bibinfo{title}{Prompt Engineering Guide}.
\newblock \bibinfo{howpublished}{\url{https://platform.openai.com/docs/guides/prompt-engineering}}.
\newblock
\newblock
\shownote{Accessed: 2024-09-01}.


\bibitem[Ou et~al\mbox{.}(2022)]%
        {ou2022indoor}
\bibfield{author}{\bibinfo{person}{Wenyan Ou}, \bibinfo{person}{Jiaming Zhang}, \bibinfo{person}{Kunyu Peng}, \bibinfo{person}{Kailun Yang}, \bibinfo{person}{Gerhard Jaworek}, \bibinfo{person}{Karin M{\"u}ller}, {and} \bibinfo{person}{Rainer Stiefelhagen}.} \bibinfo{year}{2022}\natexlab{}.
\newblock \showarticletitle{Indoor navigation assistance for visually impaired people via dynamic SLAM and panoptic segmentation with an RGB-D sensor}. In \bibinfo{booktitle}{\emph{Proceedings of International Conference on Computers Helping People with Special Needs}}. Springer, \bibinfo{pages}{160--168}.
\newblock


\bibitem[Oumard et~al\mbox{.}(2022)]%
        {oumard2022voiceui}
\bibfield{author}{\bibinfo{person}{Christina Oumard}, \bibinfo{person}{Julian Kreimeier}, {and} \bibinfo{person}{Timo G\"{o}tzelmann}.} \bibinfo{year}{2022}\natexlab{}.
\newblock \showarticletitle{Implementation and Evaluation of a Voice User Interface with Offline Speech Processing for People who are Blind or Visually Impaired}. In \bibinfo{booktitle}{\emph{Proceedings of the 15th International Conference on PErvasive Technologies Related to Assistive Environments}} (Corfu, Greece) \emph{(\bibinfo{series}{PETRA '22})}. \bibinfo{publisher}{ACM}, \bibinfo{address}{New York, NY, USA}, \bibinfo{pages}{277–285}.
\newblock
\showISBNx{9781450396318}
\urldef\tempurl%
\url{https://doi.org/10.1145/3529190.3529197}
\showDOI{\tempurl}


\bibitem[Project(nd)]%
        {ATMAPS-D2.1}
\bibfield{author}{\bibinfo{person}{ATMAPS Project}.} \bibinfo{year}{n.d.}\natexlab{}.
\newblock \bibinfo{booktitle}{\emph{User Requirements and Specifications Report (Deliverable D2.1)}}.
\newblock \bibinfo{type}{Project Deliverable}. \bibinfo{institution}{ATMAPS Consortium}.
\newblock
\urldef\tempurl%
\url{https://www.atmaps.eu/deliverables/ATMAPS-D_2_1-User_requirements_and_specifications_report.pdf}
\showURL{%
\tempurl}
\newblock
\shownote{[Online; accessed 5-April-2025]}.


\bibitem[Reimers and Gurevych(2021)]%
        {reimers2021allMiniLM}
\bibfield{author}{\bibinfo{person}{Nils Reimers} {and} \bibinfo{person}{Iryna Gurevych}.} \bibinfo{year}{2021}\natexlab{}.
\newblock \bibinfo{title}{all-MiniLM-L6-v2}.
\newblock \bibinfo{howpublished}{\url{https://huggingface.co/sentence-transformers/all-MiniLM-L6-v2}}.
\newblock
\newblock
\shownote{Accessed: 2025-04-05}.


\bibitem[Riddering(2023)]%
        {visionaware_scanning_2023}
\bibfield{author}{\bibinfo{person}{Anne Riddering}.} \bibinfo{year}{2023}\natexlab{}.
\newblock \bibinfo{title}{Scanning Efficiently for Activities of Daily Living}.
\newblock \bibinfo{howpublished}{VisionAware}.
\newblock
\urldef\tempurl%
\url{https://aphconnectcenter.org/visionaware/eye-conditions/eye-health/low-vision/scanning-efficiently-for-activities-of-daily-living/}
\showURL{%
\tempurl}
\newblock
\shownote{Accessed: 2023-09-10}.


\bibitem[Schauerte et~al\mbox{.}(2012)]%
        {schauerte2012assistive}
\bibfield{author}{\bibinfo{person}{Boris Schauerte}, \bibinfo{person}{Manel Martinez}, \bibinfo{person}{Angela Constantinescu}, {and} \bibinfo{person}{Rainer Stiefelhagen}.} \bibinfo{year}{2012}\natexlab{}.
\newblock \showarticletitle{An assistive vision system for the blind that helps find lost things}. In \bibinfo{booktitle}{\emph{Proceedings of the 13th International Conference Computers Helping People with Special Needs}}. Springer, \bibinfo{pages}{566--572}.
\newblock


\bibitem[Shinohara and Wobbrock(2016)]%
        {kristen2016self}
\bibfield{author}{\bibinfo{person}{Kristen Shinohara} {and} \bibinfo{person}{Jacob~O. Wobbrock}.} \bibinfo{year}{2016}\natexlab{}.
\newblock \showarticletitle{Self-Conscious or Self-Confident? A Diary Study Conceptualizing the Social Accessibility of Assistive Technology}.
\newblock \bibinfo{journal}{\emph{Transactions on Accessible Computing}} \bibinfo{volume}{8}, \bibinfo{number}{2}, Article \bibinfo{articleno}{5} (\bibinfo{date}{jan} \bibinfo{year}{2016}), \bibinfo{numpages}{31}~pages.
\newblock
\showISSN{1936-7228}
\urldef\tempurl%
\url{https://doi.org/10.1145/2827857}
\showDOI{\tempurl}


\bibitem[Shneiderman(1996)]%
        {shneiderman1996eyes}
\bibfield{author}{\bibinfo{person}{Ben Shneiderman}.} \bibinfo{year}{1996}\natexlab{}.
\newblock \showarticletitle{The Eyes Have It: A Task by Data Type Taxonomy for Information Visualizations}. In \bibinfo{booktitle}{\emph{Proceedings of the 1996 IEEE Symposium on Visual Languages}} \emph{(\bibinfo{series}{VL '96})}. \bibinfo{publisher}{IEEE Computer Society}, \bibinfo{address}{USA}, \bibinfo{pages}{336}.
\newblock
\showISBNx{081867508X}


\bibitem[Sugashini and Balakrishnan(2024)]%
        {Sugashini2024yologlass}
\bibfield{author}{\bibinfo{person}{T. Sugashini} {and} \bibinfo{person}{G. Balakrishnan}.} \bibinfo{year}{2024}\natexlab{}.
\newblock \showarticletitle{YOLO glass: video-based smart object detection using squeeze and attention YOLO network}.
\newblock \bibinfo{journal}{\emph{Signal, Image and Video Processing}} \bibinfo{volume}{18}, \bibinfo{number}{3} (\bibinfo{date}{4} \bibinfo{year}{2024}), \bibinfo{pages}{2105--2115}.
\newblock
\showISSN{1863-1711}
\urldef\tempurl%
\url{https://doi.org/10.1007/s11760-023-02855-x}
\showDOI{\tempurl}


\bibitem[Sun et~al\mbox{.}(2025)]%
        {sun2025objectnav}
\bibfield{author}{\bibinfo{person}{Jingwen Sun}, \bibinfo{person}{Jing Wu}, \bibinfo{person}{Ze Ji}, {and} \bibinfo{person}{Yu-Kun Lai}.} \bibinfo{year}{2025}\natexlab{}.
\newblock \showarticletitle{A Survey of Object Goal Navigation}.
\newblock \bibinfo{journal}{\emph{IEEE Transactions on Automation Science and Engineering}}  \bibinfo{volume}{22} (\bibinfo{year}{2025}), \bibinfo{pages}{2292--2308}.
\newblock
\urldef\tempurl%
\url{https://doi.org/10.1109/TASE.2024.3378010}
\showDOI{\tempurl}


\bibitem[Surougi and McCann(2023)]%
        {surougi2023realtime_path_planning}
\bibfield{author}{\bibinfo{person}{Hadeel~R. Surougi} {and} \bibinfo{person}{Julie~A. McCann}.} \bibinfo{year}{2023}\natexlab{}.
\newblock \showarticletitle{Real-Time Optimisation-Based Path Planning for Visually Impaired People in Dynamic Environments}. In \bibinfo{booktitle}{\emph{Proceedings of the IEEE/CVF International Conference on Computer Vision Workshops (ICCVW)}}. IEEE, \bibinfo{pages}{1831--1840}.
\newblock


\bibitem[Tahoun et~al\mbox{.}(2020)]%
        {Nourhan2020smart}
\bibfield{author}{\bibinfo{person}{Nourhan Tahoun}, \bibinfo{person}{Anwar Awad}, {and} \bibinfo{person}{Talal Bonny}.} \bibinfo{year}{2020}\natexlab{}.
\newblock \showarticletitle{Smart Assistant for Blind and Visually Impaired People}. In \bibinfo{booktitle}{\emph{Proceedings of the 3rd International Conference on Advances in Artificial Intelligence}} (Istanbul, Turkey) \emph{(\bibinfo{series}{ICAAI '19})}. \bibinfo{publisher}{ACM}, \bibinfo{address}{New York, NY, USA}, \bibinfo{pages}{227–231}.
\newblock
\showISBNx{9781450372534}
\urldef\tempurl%
\url{https://doi.org/10.1145/3369114.3369139}
\showDOI{\tempurl}


\bibitem[Taioli et~al\mbox{.}(2024)]%
        {taioli2024collaborative}
\bibfield{author}{\bibinfo{person}{Francesco Taioli}, \bibinfo{person}{Edoardo Zorzi}, \bibinfo{person}{Gianni Franchi}, \bibinfo{person}{Alberto Castellini}, \bibinfo{person}{Alessandro Farinelli}, \bibinfo{person}{Marco Cristani}, {and} \bibinfo{person}{Yiming Wang}.} \bibinfo{year}{2024}\natexlab{}.
\newblock \showarticletitle{Collaborative Instance Navigation: Leveraging Agent Self-Dialogue to Minimize User Input}.
\newblock \bibinfo{journal}{\emph{arXiv preprint arXiv:2412.01250}} (\bibinfo{year}{2024}).
\newblock


\bibitem[{TapTapSee, Inc.}(nd)]%
        {taptapsee}
\bibfield{author}{\bibinfo{person}{{TapTapSee, Inc.}}} \bibinfo{year}{n.d.}\natexlab{}.
\newblock \bibinfo{title}{TapTapSee}.
\newblock
\newblock
\urldef\tempurl%
\url{https://www.taptapseeapp.com}
\showURL{%
\tempurl}
\newblock
\shownote{Version 3.1.1 [Mobile application software]}.


\bibitem[V7(2024)]%
        {ai_poly}
\bibfield{author}{\bibinfo{person}{V7}.} \bibinfo{year}{2024}\natexlab{}.
\newblock \bibinfo{title}{Aipoly}.
\newblock \bibinfo{howpublished}{\url{https://www.aipoly.com}}.
\newblock
\newblock
\shownote{Accessed: 2024-07-04}.


\bibitem[Vincenzi(2021)]%
        {beatrice2021guiding}
\bibfield{author}{\bibinfo{person}{Beatrice Vincenzi}.} \bibinfo{year}{2021}\natexlab{}.
\newblock \showarticletitle{AI assistive technology for extending sighted guiding}.
\newblock \bibinfo{journal}{\emph{SIGACCESS Access. Comput.}} \bibinfo{number}{129}, Article \bibinfo{articleno}{7} (\bibinfo{date}{mar} \bibinfo{year}{2021}), \bibinfo{numpages}{5}~pages.
\newblock
\showISSN{1558-2337}
\urldef\tempurl%
\url{https://doi.org/10.1145/3458055.3458062}
\showDOI{\tempurl}


\bibitem[Waisberg et~al\mbox{.}(2024)]%
        {Waisberg2024}
\bibfield{author}{\bibinfo{person}{Ethan Waisberg}, \bibinfo{person}{Joshua Ong}, \bibinfo{person}{Mouayad Masalkhi}, \bibinfo{person}{Nasif Zaman}, \bibinfo{person}{Prithul Sarker}, \bibinfo{person}{Andrew~G. Lee}, {and} \bibinfo{person}{Alireza Tavakkoli}.} \bibinfo{year}{2024}\natexlab{}.
\newblock \showarticletitle{Meta smart glasses—large language models and the future for assistive glasses for individuals with vision impairments}.
\newblock \bibinfo{journal}{\emph{Eye}} \bibinfo{volume}{38}, \bibinfo{number}{6} (\bibinfo{year}{2024}), \bibinfo{pages}{1036--1038}.
\newblock
\showISSN{1476-5454}
\urldef\tempurl%
\url{https://doi.org/10.1038/s41433-023-02842-z}
\showDOI{\tempurl}


\bibitem[Wang et~al\mbox{.}(2024)]%
        {wang2024visiongpt}
\bibfield{author}{\bibinfo{person}{Hao Wang}, \bibinfo{person}{Jiayou Qin}, \bibinfo{person}{Ashish Bastola}, \bibinfo{person}{Xiwen Chen}, \bibinfo{person}{John Suchanek}, \bibinfo{person}{Zihao Gong}, {and} \bibinfo{person}{Abolfazl Razi}.} \bibinfo{year}{2024}\natexlab{}.
\newblock \showarticletitle{VisionGPT: LLM-Assisted Real-Time Anomaly Detection for Safe Visual Navigation}.
\newblock \bibinfo{journal}{\emph{arXiv preprint arXiv:2403.12415}} (\bibinfo{year}{2024}).
\newblock


\bibitem[Wen et~al\mbox{.}(2024)]%
        {wen2024findmythings}
\bibfield{author}{\bibinfo{person}{Linda~Yilin Wen}, \bibinfo{person}{Cecily Morrison}, \bibinfo{person}{Martin Grayson}, \bibinfo{person}{Rita~Faia Marques}, \bibinfo{person}{Daniela Massiceti}, \bibinfo{person}{Camilla Longden}, {and} \bibinfo{person}{Edward Cutrell}.} \bibinfo{year}{2024}\natexlab{}.
\newblock \showarticletitle{Find My Things: Personalized Accessibility through Teachable AI for People who are Blind or Low Vision}. In \bibinfo{booktitle}{\emph{Extended Abstracts of the CHI Conference on Human Factors in Computing Systems}} (Honolulu, HI, USA) \emph{(\bibinfo{series}{CHI EA '24})}. \bibinfo{publisher}{Association for Computing Machinery}, \bibinfo{address}{New York, NY, USA}, Article \bibinfo{articleno}{403}, \bibinfo{numpages}{6}~pages.
\newblock
\showISBNx{9798400703317}
\urldef\tempurl%
\url{https://doi.org/10.1145/3613905.3648641}
\showDOI{\tempurl}


\bibitem[{World Health Organization}(2021)]%
        {WHO2021ICD10}
\bibfield{author}{\bibinfo{person}{{World Health Organization}}.} \bibinfo{year}{2021}\natexlab{}.
\newblock \bibinfo{title}{ICD-10: International Statistical Classification of Diseases and Related Health Problems, 10th Revision}.
\newblock \bibinfo{howpublished}{\url{https://icd.who.int/browse10/2021/en/H54}}.
\newblock
\newblock
\shownote{Accessed: 2023-08-05}.


\bibitem[Xie et~al\mbox{.}(2022a)]%
        {xie2022iterative}
\bibfield{author}{\bibinfo{person}{Jingyi Xie}, \bibinfo{person}{Madison Reddie}, \bibinfo{person}{Sooyeon Lee}, \bibinfo{person}{Syed~Masum Billah}, \bibinfo{person}{Zihan Zhou}, \bibinfo{person}{Chun-Hua Tsai}, {and} \bibinfo{person}{John~M. Carroll}.} \bibinfo{year}{2022}\natexlab{a}.
\newblock \showarticletitle{Iterative Design and Prototyping of Computer Vision Mediated Remote Sighted Assistance}.
\newblock \bibinfo{journal}{\emph{ACM Trans. Comput.-Hum. Interact.}} \bibinfo{volume}{29}, \bibinfo{number}{4}, Article \bibinfo{articleno}{36} (\bibinfo{date}{March} \bibinfo{year}{2022}), \bibinfo{numpages}{40}~pages.
\newblock
\showISSN{1073-0516}
\urldef\tempurl%
\url{https://doi.org/10.1145/3501298}
\showDOI{\tempurl}


\bibitem[Xie et~al\mbox{.}(2022b)]%
        {xie2022help}
\bibfield{author}{\bibinfo{person}{Jingyi Xie}, \bibinfo{person}{Rui Yu}, \bibinfo{person}{Sooyeon Lee}, \bibinfo{person}{Yao Lyu}, \bibinfo{person}{Syed~Masum Billah}, {and} \bibinfo{person}{John~M. Carroll}.} \bibinfo{year}{2022}\natexlab{b}.
\newblock \showarticletitle{Helping Helpers: Supporting Volunteers in Remote Sighted Assistance with Augmented Reality Maps}. In \bibinfo{booktitle}{\emph{Proceedings of the 2022 ACM Designing Interactive Systems Conference}} (Virtual Event, Australia) \emph{(\bibinfo{series}{DIS '22})}. \bibinfo{publisher}{Association for Computing Machinery}, \bibinfo{address}{New York, NY, USA}, \bibinfo{pages}{881–897}.
\newblock
\showISBNx{9781450393584}
\urldef\tempurl%
\url{https://doi.org/10.1145/3532106.3533560}
\showDOI{\tempurl}


\bibitem[Xie et~al\mbox{.}(2025)]%
        {xie2025beyond}
\bibfield{author}{\bibinfo{person}{Jingyi Xie}, \bibinfo{person}{Rui Yu}, \bibinfo{person}{He Zhang}, \bibinfo{person}{Syed~Masum Billah}, \bibinfo{person}{Sooyeon Lee}, {and} \bibinfo{person}{John~M Carroll}.} \bibinfo{year}{2025}\natexlab{}.
\newblock \showarticletitle{Beyond Visual Perception: Insights from Smartphone Interaction of Visually Impaired Users with Large Multimodal Models}. In \bibinfo{booktitle}{\emph{Proceedings of the 2025 CHI Conference on Human Factors in Computing Systems}}.
\newblock


\bibitem[Xie et~al\mbox{.}(2024)]%
        {xie2024emerging}
\bibfield{author}{\bibinfo{person}{Jingyi Xie}, \bibinfo{person}{Rui Yu}, \bibinfo{person}{He Zhang}, \bibinfo{person}{Sooyeon Lee}, \bibinfo{person}{Syed~Masum Billah}, {and} \bibinfo{person}{John~M Carroll}.} \bibinfo{year}{2024}\natexlab{}.
\newblock \showarticletitle{Emerging practices for large multimodal model (lmm) assistance for people with visual impairments: Implications for design}.
\newblock \bibinfo{journal}{\emph{arXiv preprint arXiv:2407.08882}} (\bibinfo{year}{2024}).
\newblock


\bibitem[Yang et~al\mbox{.}(2024)]%
        {yang2024viassist}
\bibfield{author}{\bibinfo{person}{Bufang Yang}, \bibinfo{person}{Lixing He}, \bibinfo{person}{Kaiwei Liu}, {and} \bibinfo{person}{Zhenyu Yan}.} \bibinfo{year}{2024}\natexlab{}.
\newblock \showarticletitle{VIAssist: Adapting Multi-Modal Large Language Models for Users with Visual Impairments}. In \bibinfo{booktitle}{\emph{Proceedings of the IEEE International Workshop on Foundation Models for Cyber-Physical Systems \& Internet of Things (FMSys)}}. \bibinfo{pages}{32--37}.
\newblock
\urldef\tempurl%
\url{https://doi.org/10.1109/FMSys62467.2024.00010}
\showDOI{\tempurl}


\bibitem[Yang et~al\mbox{.}(2018a)]%
        {yang2018unifying}
\bibfield{author}{\bibinfo{person}{Kailun Yang}, \bibinfo{person}{Luis~M Bergasa}, \bibinfo{person}{Eduardo Romera}, \bibinfo{person}{Ruiqi Cheng}, \bibinfo{person}{Tianxue Chen}, {and} \bibinfo{person}{Kaiwei Wang}.} \bibinfo{year}{2018}\natexlab{a}.
\newblock \showarticletitle{Unifying terrain awareness through real-time semantic segmentation}. In \bibinfo{booktitle}{\emph{Proceedings of the IEEE Intelligent Vehicles Symposium (IV)}}. IEEE, \bibinfo{pages}{1033--1038}.
\newblock


\bibitem[Yang et~al\mbox{.}(2018b)]%
        {yang2018predicting}
\bibfield{author}{\bibinfo{person}{Kailun Yang}, \bibinfo{person}{Luis~M Bergasa}, \bibinfo{person}{Eduardo Romera}, \bibinfo{person}{Xiao Huang}, {and} \bibinfo{person}{Kaiwei Wang}.} \bibinfo{year}{2018}\natexlab{b}.
\newblock \showarticletitle{Predicting polarization beyond semantics for wearable robotics}. In \bibinfo{booktitle}{\emph{Proceedings of the IEEE-RAS 18th International Conference on Humanoid Robots (Humanoids)}}. IEEE, \bibinfo{pages}{96--103}.
\newblock


\bibitem[Yang et~al\mbox{.}(2010)]%
        {yang2010context}
\bibfield{author}{\bibinfo{person}{Xiaodong Yang}, \bibinfo{person}{YingLi Tian}, \bibinfo{person}{Chucai Yi}, {and} \bibinfo{person}{Aries Arditi}.} \bibinfo{year}{2010}\natexlab{}.
\newblock \showarticletitle{Context-based indoor object detection as an aid to blind persons accessing unfamiliar environments}. In \bibinfo{booktitle}{\emph{ACMMM}} (Firenze, Italy) \emph{(\bibinfo{series}{MM '10})}. \bibinfo{publisher}{Association for Computing Machinery}, \bibinfo{address}{New York, NY, USA}, \bibinfo{pages}{1087–1090}.
\newblock
\showISBNx{9781605589336}
\urldef\tempurl%
\url{https://doi.org/10.1145/1873951.1874156}
\showDOI{\tempurl}


\bibitem[Yi et~al\mbox{.}(2013)]%
        {Yi2013}
\bibfield{author}{\bibinfo{person}{Chucai Yi}, \bibinfo{person}{Roberto~W. Flores}, \bibinfo{person}{Ricardo Chincha}, {and} \bibinfo{person}{YingLi Tian}.} \bibinfo{year}{2013}\natexlab{}.
\newblock \showarticletitle{Finding objects for assisting blind people}.
\newblock \bibinfo{journal}{\emph{Network Modeling Analysis in Health Informatics and Bioinformatics}} \bibinfo{volume}{2}, \bibinfo{number}{2} (\bibinfo{year}{2013}), \bibinfo{pages}{71--79}.
\newblock
\showISSN{2192-6670}
\urldef\tempurl%
\url{https://doi.org/10.1007/s13721-013-0026-x}
\showDOI{\tempurl}


\bibitem[Yin et~al\mbox{.}(2024)]%
        {yin2024sgnav}
\bibfield{author}{\bibinfo{person}{Hang Yin}, \bibinfo{person}{Xiuwei Xu}, \bibinfo{person}{Zhenyu Wu}, \bibinfo{person}{Jie Zhou}, {and} \bibinfo{person}{Jiwen Lu}.} \bibinfo{year}{2024}\natexlab{}.
\newblock \showarticletitle{SG-Nav: Online 3D Scene Graph Prompting for LLM-based Zero-shot Object Navigation}. In \bibinfo{booktitle}{\emph{Advances in Neural Information Processing Systems (NeurIPS)}}.
\newblock


\bibitem[Yin et~al\mbox{.}(2025)]%
        {yin2025unigoal}
\bibfield{author}{\bibinfo{person}{Hang Yin}, \bibinfo{person}{Xiuwei Xu}, \bibinfo{person}{Lingqing Zhao}, \bibinfo{person}{Ziwei Wang}, \bibinfo{person}{Jie Zhou}, {and} \bibinfo{person}{Jiwen Lu}.} \bibinfo{year}{2025}\natexlab{}.
\newblock \showarticletitle{UniGoal: Towards Universal Zero-shot Goal-oriented Navigation}. In \bibinfo{booktitle}{\emph{CVPR}}.
\newblock


\bibitem[Yokoyama et~al\mbox{.}(2024)]%
        {yokoyama2024hm3d}
\bibfield{author}{\bibinfo{person}{Naoki Yokoyama}, \bibinfo{person}{Ram Ramrakhya}, \bibinfo{person}{Abhishek Das}, \bibinfo{person}{Dhruv Batra}, {and} \bibinfo{person}{Sehoon Ha}.} \bibinfo{year}{2024}\natexlab{}.
\newblock \showarticletitle{HM3D-OVON: A dataset and benchmark for open-vocabulary object goal navigation}. In \bibinfo{booktitle}{\emph{2024 IEEE/RSJ International Conference on Intelligent Robots and Systems (IROS)}}. IEEE, \bibinfo{pages}{5543--5550}.
\newblock


\bibitem[Zhang et~al\mbox{.}(2023)]%
        {mobile_sam}
\bibfield{author}{\bibinfo{person}{Chaoning Zhang}, \bibinfo{person}{Dongshen Han}, \bibinfo{person}{Yu Qiao}, \bibinfo{person}{Jung~Uk Kim}, \bibinfo{person}{Sung-Ho Bae}, \bibinfo{person}{Seungkyu Lee}, {and} \bibinfo{person}{Choong~Seon Hong}.} \bibinfo{year}{2023}\natexlab{}.
\newblock \showarticletitle{Faster Segment Anything: Towards Lightweight SAM for Mobile Applications}.
\newblock \bibinfo{journal}{\emph{arXiv preprint arXiv:2306.14289}} (\bibinfo{year}{2023}).
\newblock


\bibitem[Zhang et~al\mbox{.}(2025)]%
        {zhang2025navigpt}
\bibfield{author}{\bibinfo{person}{He Zhang}, \bibinfo{person}{Nicholas~J. Falletta}, \bibinfo{person}{Jingyi Xie}, \bibinfo{person}{Rui Yu}, \bibinfo{person}{Sooyeon Lee}, \bibinfo{person}{Syed~Masum Billah}, {and} \bibinfo{person}{John~M. Carroll}.} \bibinfo{year}{2025}\natexlab{}.
\newblock \showarticletitle{Enhancing the Travel Experience for People with Visual Impairments through Multimodal Interaction: NaviGPT, A Real-Time AI-Driven Mobile Navigation System}. In \bibinfo{booktitle}{\emph{Companion Proceedings of the 2025 ACM International Conference on Supporting Group Work}} (Hilton Head, New Jersey, USA) \emph{(\bibinfo{series}{GROUP '25})}. \bibinfo{publisher}{Association for Computing Machinery}, \bibinfo{address}{New York, NY, USA}, \bibinfo{pages}{29–35}.
\newblock
\showISBNx{9798400711879}
\urldef\tempurl%
\url{https://doi.org/10.1145/3688828.3699636}
\showDOI{\tempurl}


\bibitem[Zhang et~al\mbox{.}(2021)]%
        {zhang2021trans4trans}
\bibfield{author}{\bibinfo{person}{Jiaming Zhang}, \bibinfo{person}{Kailun Yang}, \bibinfo{person}{Angela Constantinescu}, \bibinfo{person}{Kunyu Peng}, \bibinfo{person}{Karin M{\"u}ller}, {and} \bibinfo{person}{Rainer Stiefelhagen}.} \bibinfo{year}{2021}\natexlab{}.
\newblock \showarticletitle{Trans4Trans: Efficient transformer for transparent object segmentation to help visually impaired people navigate in the real world}. In \bibinfo{booktitle}{\emph{Proceedings of the IEEE/CVF International Conference on Computer Vision Workshops (ICCVW)}}. IEEE, \bibinfo{pages}{1760--1770}.
\newblock


\bibitem[Zhao et~al\mbox{.}(2024)]%
        {zhao2024vialm}
\bibfield{author}{\bibinfo{person}{Yi Zhao}, \bibinfo{person}{Yilin Zhang}, \bibinfo{person}{Rong Xiang}, \bibinfo{person}{Jing Li}, {and} \bibinfo{person}{Hillming Li}.} \bibinfo{year}{2024}\natexlab{}.
\newblock \showarticletitle{VIALM: A Survey and Benchmark of Visually Impaired Assistance with Large Models}.
\newblock \bibinfo{journal}{\emph{arXiv preprint arXiv:2402.01735}} (\bibinfo{year}{2024}).
\newblock


\bibitem[Zheng et~al\mbox{.}(2024)]%
        {zheng2023materobot}
\bibfield{author}{\bibinfo{person}{Junwei Zheng}, \bibinfo{person}{Jiaming Zhang}, \bibinfo{person}{Kailun Yang}, \bibinfo{person}{Kunyu Peng}, {and} \bibinfo{person}{Rainer Stiefelhagen}.} \bibinfo{year}{2024}\natexlab{}.
\newblock \showarticletitle{MateRobot: Material Recognition in Wearable Robotics for People with Visual Impairments}. In \bibinfo{booktitle}{\emph{Proceedings of the IEEE International Conference on Robotics and Automation (ICRA)}}. IEEE, \bibinfo{pages}{2303--2309}.
\newblock


\bibitem[Zou et~al\mbox{.}(2023)]%
        {zou2023realtime_passable}
\bibfield{author}{\bibinfo{person}{Wenbin Zou}, \bibinfo{person}{Guoguang Hua}, \bibinfo{person}{Yue Zhuang}, {and} \bibinfo{person}{Shishun Tian}.} \bibinfo{year}{2023}\natexlab{}.
\newblock \showarticletitle{Real-time passable area segmentation with consumer RGB-D cameras for the visually impaired}.
\newblock \bibinfo{journal}{\emph{IEEE Transactions on Instrumentation and Measurement}}  \bibinfo{volume}{72} (\bibinfo{year}{2023}), \bibinfo{pages}{1--11}.
\newblock


\end{thebibliography}
\appendix

\begin{algorithm*}[!ht]
  \caption{Pseudo code for instruction data generation with GPT-4V~\cite{openai-gpt4}.}
  \label{alg:gpt4v}
\var{PROMPT\_DICT}\{ \\
    \KwSty{prompt\_system}: 
    (\\ \hspace{2mm}``You are an AI visual assistant, observing scenarios from the egocentric perspective of a user who is blind or visually impaired. The user will present various prompts regarding scene description, route planning, and open-ended questions. Responses should be concise and practical, not exceeding 100 words in length. Ensure that your tone reflects that of a visual AI assistant interpreting and responding to the scene. Craft your responses with consideration of the following perspectives: position, count, size, color, material, and shape.''
    ), \\ 
    \KwSty{prompt\_route\_planning}: 
    (\\ \hspace{2mm}``I am a blind person. Please guide me on how to approach this \{target\_object\} based on this picture. At the beginning of your response, always remind me to align my body with my head's direction.'' ), \\
    \KwSty{prompt\_scene\_description}: 
    (\\ \hspace{2mm}``Please describe the scene. You need to provide the positional relationship between the items, and your answer should be brief.'' )\}
 
\KwSty{output} = \FuncCall{openai.ChatCompletion.create}{ \\
\hspace{13mm}     \var{model=``gpt-4v''}, \\
\hspace{13mm}  \var{messages=[
\{``role'': ``system'', ``content'': \KwSty{prompt\_system} \},
\\ 
\hspace{36mm}\{``role'': ``user'', ``content'': \KwSty{Image}; \KwSty{prompt\_function}} \}]
    }
\end{algorithm*}
\end{document}